\documentclass[%
 reprint,
 amsmath,amssymb,
 aps,
]{revtex4-2}

\usepackage{graphicx}
\usepackage{dcolumn}
\usepackage{bm}
\usepackage{braket}
\usepackage{xcolor}
\usepackage{comment}
\usepackage[english]{babel}

\usepackage{algorithm}
\usepackage{algpseudocode}

\algrenewcommand\algorithmicrequire{\textbf{Input:}}
\algrenewcommand\algorithmicensure{\textbf{Output:}}
\usepackage{float}

\begin{document}

\title{Decoy state and purification protocols for superior quantum key distribution with imperfect quantum-dot based single photon sources: Theory and Experiment}

\author{Yuval Bloom$^{*}$}
\author{Yoad Ordan}%
\thanks{These authors contributed equally to this work}
\author{Tamar Levin}
\author{Kfir Sulimany}%
\author{Ronen Rapaport}
\email{ronen.rapaport@mail.huji.ac.il}
\affiliation{%
Racah Institute of Physics, The Hebrew University of Jerusalem, Jerusalem 9190401, Israel
}%

\author{Eric G. Bowes}
\author{Jennifer A. Hollingsworth}
\affiliation{Materials Physics \& Applications Division: Center for Integrated Nanotechnologies, Los Alamos National Laboratory, Los Alamos, New Mexico 87545, USA}

\date{\today}

\begin{abstract}
The original proposal of quantum key distribution (QKD) was based on ideal single photon sources, which 40 years later, are still challenging to develop. Therefore, the development of decoy state protocols using weak coherent states (WCS) from lasers, set the frontier in terms of secure key rates and distances. Here, we propose and experimentally demonstrate two simple-to-implement QKD protocols that allow practical, far from ideal sub-Poissonian photon sources to outperform state-of-the-art WCS. By engineering the photon statistics of a biexciton-exciton cascade in room temperature single photon sources based on giant colloidal quantum dots coupled to nanoantennas, we show that either a truncated decoy state protocol or a heralded purification protocol can be employed to achieve a significantly increased performance in terms of the maximal allowed channel loss for secure key creation, which can exceed even that of ideal WCS by more than 3dB. We then experimentally emulate a BB84 QKD using such a quantum dot source, verifying the superiority of our protocols over the best possible BB84 WCS performance. These protocols can be utilized efficiently on a host of various quantum emitters having controllable photon statistics with a finite photon-number basis, offering a practical approach to QKD without the hindering requirements on the single photon purity of the photon source.

\end{abstract}

\maketitle

\section{\label{sec:intro}Introduction}

Quantum key distribution (QKD) stands as a prominent candidate for the post-quantum-computer optical communication, enabling a secure key-establishing protocol between two parties employing the quantum nature of light \cite{Shor2000SimpleProtocol,Alleaume2014UsingSurvey,Pirandola2020AdvancesCryptography}. In theory, encoding the information for QKD onto pure single photon states is ultimately secured. In practice, it is limited by imperfections of the transmitter and the receiver, making security proofs a challenging task \cite{Yuen2012,Lo2014SecureDistribution}, thus reducing the maximal secure key rate (SKR) and the maximal losses in the communication channel over which SKR is achievable \cite{Xu2020SecureDevices, Scarani2009TheDistribution}.

One of the main practical limitations of QKD systems is the lack of an ideal source for single photons (SPS), required to prevent eavesdropping attacks such as photon number splitting attack (PNS) \cite{Pousa2024ComparisonDistribution}. The SKR and maximal allowed channel loss (MCL) of simple QKD protocols decrease dramatically with the increase of two-or-more photon events being transmitted from the source \cite{Somma2013SecurityAttacks, Pereira2019QuantumSources}. This leads to very stringent requirements on both the single photon purity and the photon emission rate of any practical SPS \cite{Lo2005DecoyDistribution}.

Despite many years of exploration and engineering \cite{Eisaman2011InvitedDetectors, Lounis2005Single-photonSources}, a simple, stable SPS system that has the required aforementioned properties and can be employed in real-life QKD applications is still an outstanding challenge, which led researchers to suggest and utilize protocols employing attenuated lasers emitting weak coherent states (WCS) \cite{stucki2005fast}. As WCS has a Poissonian distribution of photons, and thus a two-or-more photons probability is never vanishing, advanced protocols such as decoy state protocols have been developed to enhance both SKR and MCL \cite{Lo2005DecoyDistribution,Ma2005PracticalDistribution,islam2017provably,liao2017satellite,sulimany2021high}.

Decoy state protocols introduce variations in the distribution of the emitted photons to better characterize the channel and obtain a tighter bound on the SKR \cite{Zhang2015ApproachingStates,Sun2013PracticalDistribution, Lim2014ConciseDistribution, Kraus2007SecurityPulses}, allowing detection of PNS attacks \cite{Wang2005BeatingCryptography, Fan-Yuan2021OptimizingSystems}.
Yet, even with decoy state protocols, WCS are highly sensitive \cite{Chen2022ExperimentalImperfections, Grice2019QuantumStates} to the high probability of the vacuum state, due to the dependence between the different photon number emission probabilities, which overall limits the possible SKR \cite{Somma2013SecurityAttacks}. The above solutions and inherent limitations of WCS for QKD set clear opportunities for improving the security and efficiency of QKD applications using realistic SPS systems, such as solid-state emitters \cite{Zhang2024ExperimentalLimit, somaschi2016near, uppu2020scalable, muller2017quantum, loredo2016scalable, he2013demand} and spontaneous parametric down conversion heralded sources \cite{zhan2023quantum,zhan2025experimental}. Yet, while efforts to develop a nearly ideal SPS are still ongoing by many groups \cite{Bozzio2022EnhancingSources, Kagan2021ColloidalScience, Kim2023QuantumTechnology, Qian2012EngineeredSources, Nelson2024ColloidalSources}, they still fall short as compared to the performance of QKD protocols based on WCS incorporating decoy states.

Here we introduce a radically different, more realistic approach. By utilizing two new simple QKD protocols on compact, imperfect SPS sources, we significantly enhance the QKD performance, in terms of the MCL, as compared to even ideal, infinite decoy state protocols using WCS.

We base our protocols on our recent demonstrations of room temperature operating SPS devices based on a giant colloidal nanocrystal quantum dot (gCQD) coupled to a hybrid nanoantenna and plasmonic nanocone \cite{Bloom2024RadialPol, Abudayyeh2021OvercomingSource, Lubotzky2024Room-TemperatureNanoantennas, Abudayyeh2021SingleNanoantennas,halevi2024high}. These SPS devices displayed a highly enhanced rate of photon emission approaching the GHz range, together with near-unity collection efficiencies ($>90\%$) of the emitted photons. The photon emission from such gCQDs is based on the biexciton - exciton (BX-X) emission cascade under optical excitation \cite{Li2020Biexciton}, and has been shown to be well described by a truncated photon-number Fock space with $N=0,1$, and $2$ photons, with a negligible probability of $N>2$ photons \cite{Abudayyeh2019PurificationSources}.

The two protocols are both based on our ability to vary and control the probabilities of emitting the different Fock states, $\{ P_{0}, P_{1}, P_{2} \}$ over a wide range of values, by simply varying the optical excitation power. The first protocol we implement is a simple and realistic decoy state protocol for such a source. We show that with this new protocol, we can significantly exceed, by more than $3$ dB, the MCL value of an ideal decoy state protocol using a WCS source. The second protocol uses an alternative approach based on heralded photon purification under saturated excitation.  We show that a similar performance enhancement can also be achieved. 
Later, we experimentally verify that the probability for $N>2$ ($P_{N>2}$) is negligibly small, justifying our protocols assumptions. 
Finally, we experimentally fully emulate a free-space BB84 QKD using a bare gCQD source without an antenna. The results agree well with the models, and demonstrate the superiority of our protocols over the best possible BB84 WCS performance even for such a simple imperfect SPS source. This shows that even existing SPS, which are far from being ideal, can outperform the current state-of-the-art protocols based on WCS sources.

\begin{figure*}[!thb]
    \includegraphics[scale=0.75]{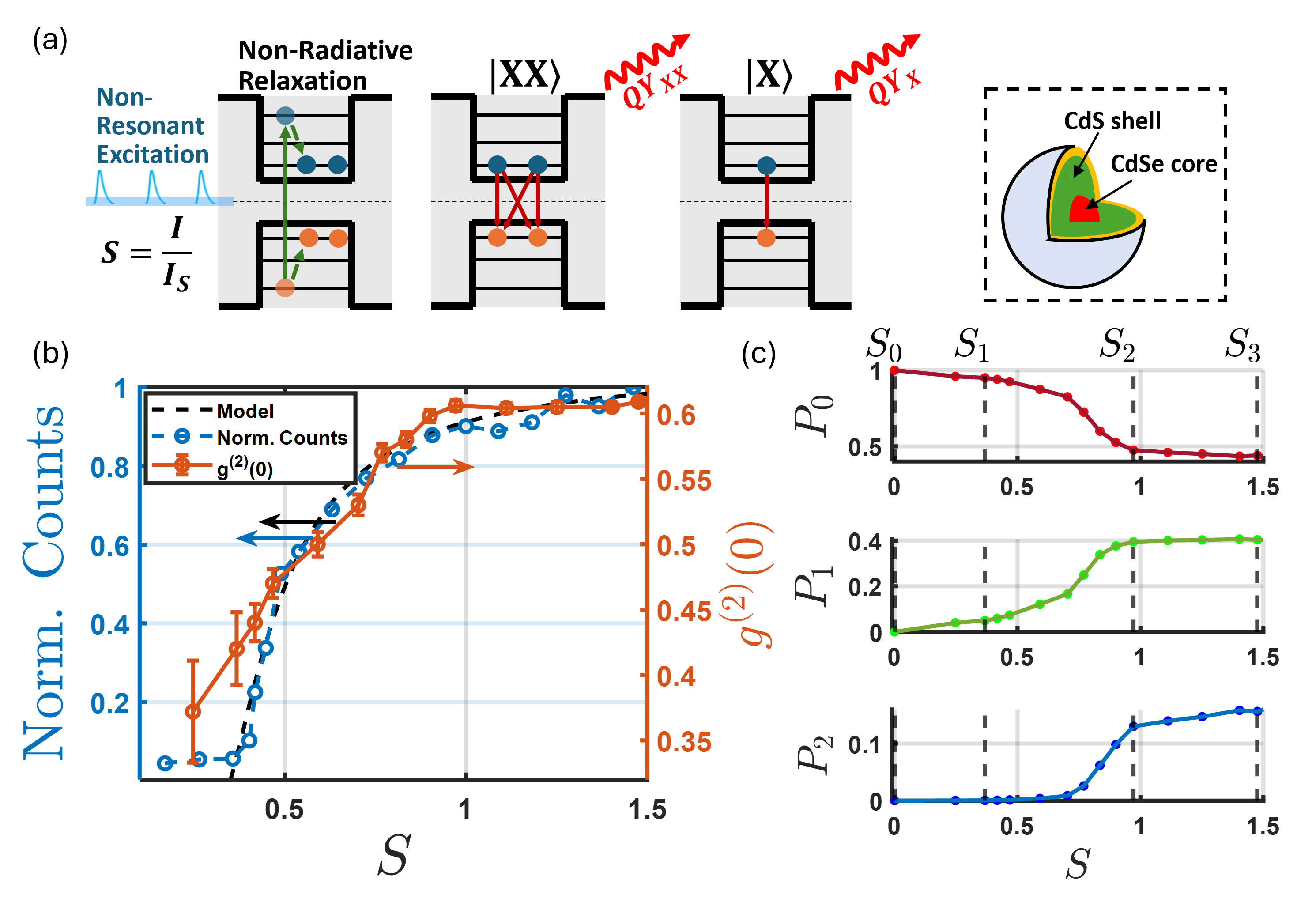}
    \caption{(a) Schematic of the BX-X cascade in an optically excited gCQD, using a non-resonant pulsed excitation with a normalized intensity $S=\frac{I}{I_S}$. Inset is a sketch of a typical gCQD, consisting of a CdSe core and a CdS shell. 
    (b) Normalized emission counts (blue open circles) as a function of $S$, displaying a saturation behaviour. The dashed black line is a fit to a model, detailed in Section II. The orange circles show the experimentally extracted $g^{(2)}(0)$ for the different powers. (c) Photon probabilities as a function of $S$. The dashed black lines mark the intensities chosen for the signal $S_2$, and the two decoy states $S_0$, $S_1$, for the decoy state on a truncated basis (DTB) protocol, and the intensity $S_{3}$, chosen for the heralded purification (HP) protocol.} 
    
    \label{fig:fig1}
\end{figure*}

\section{\label{sec:section1} Optical control of photon statistics from the BX-X emission cascade in a \lowercase{g}CQD}

Fig. \ref{fig:fig1}(a) presents a schematic sketch of the BX-X cascade in a gCQD (see Appendix A) following a non-resonant pulse excitation with power $I$. Modelling this process, a BX state $\ket{XX}$ is excited by the absorption of \textit{two pump photons}. This process happens with a probability:
\begin{equation} \label{eq: biexciton prob}
    P_{XX}=\frac{\left( \alpha I \right)^2}{1+\alpha I +\left( \alpha I\right)^2}.
\end{equation}
The probability of excitation of only the X state, $\ket{X}$, by the absorption of \textit{only one pump photon}, is given by:
\begin{equation} \label{eq: exciton prob}
    P_{X}=\frac{\alpha I}{1+\alpha I +\left( \alpha I\right)^2}.
\end{equation}
Here $\alpha$ is a constant related to the absorption cross-section of the gCQD  and we assume that any larger excitonic complexes are negligible \cite{Abudayyeh2019PurificationSources}, as will be verified later in this paper.

Following an optical excitation, the biexciton state $\ket{XX}$ decays into a single exciton state $\ket{X}$ either radiatively, by emitting a single photon with a probability $QY_{XX}$, representing the BX quantum yield, or non-radiatively with a probability $1-QY_{XX}$. The exciton state $\ket{X}$ can then recombine radiatively (non-radiatively) with probabilities $QY_{X}$, ($1-QY_X$), to the ground state such that the probability to emit zero, one or two photons, defined as $ \{ P_0, \, P_1, \, P_2 \}$ respectively, is:
\begin{align}
    P_{1}  &= P_{XX} \left( QY_{X}+QY_{XX}-2QY_{X}QY_{XX} \right) \\
           &+P_{X}QY_{X} \nonumber \\
    P_{2}  &= P_{XX}QY_{X}QY_{XX} \\
    P_{0}  &= 1-P_{1}-P_{2}
\end{align}
Since $P_{XX},P_X$ depend on the excitation power, the probabilities to emit two, one or zero photons following excitation become excitation power dependent, allowing an external control over the emitted photon statistics as is required for decoy states protocols. Such a saturable behavior of the excitonic population allows us to define a relative excitation power $S=I/I_{S}$, where $I_{S}$ is the saturation power, for which the detected count rate (which is linearly proportional to the average emission photon number) reaches $90\%$ (a convenient arbitrary value) of the maximum value (which asymptotically occurs at $I\rightarrow\infty$).

In Fig. \ref{fig:fig1}(b), we experimentally demonstrate photon statistics control by varying the excitation power from a single bare gCQD. As $S$ increases, the collected  photon emission intensity increases sublinearly and saturates at $S\gtrsim 1$. A good fit of the theoretical model, based on the absorption probabilities given in Eqs. \ref{eq: biexciton prob}, \ref{eq: exciton prob} and on the subsequent emission probabilities  $QY_{X},QY_{XX}$ is shown (see Appendix B for details). This fit confirms the basic assumptions given above for the BX-X cascade process.

At the same time, the measured second-order correlation function, $g^{(2)}(0)$, increases in a similar manner. This is understood in the following way: at low powers, most emission events are empty as $P_0$ is the dominant term. As $S$ increases, $P_1$ increases linearly (Eq. \ref{eq: exciton prob}), while $P_2$ increases quadratically and thus is negligible at $S\ll 1$. Since $g^{(2)}(0)$ is an increasing function of $P_2/(\sum_{i=0,1,2} {P_i})$, a saturable behavior of $g^{(2)}(0)$ with increasing $S$ is also expected.

Importantly, as we also show experimentally (see Appendix D), multi-excitonic ($N>2$) emissions from the gCQD can be neglected due to their low emission probability, thus $P_{N>2}=0$.

By measuring both $g^{(2)}(0)(S)$ and the count-rate $C(S)$, we can extract the probability distribution precisely, as is detailed in Appendix C.

In Fig. \ref{fig:fig1}(c), we present the extracted photon emission probabilities  $\{ P_{0}, P_{1}, P_{2} \}(S)$. As expected, the probability of emitting one or two photons increases with excitation power, but at a different rate, allowing optical control of the emitted photon statistics with only varying $S$.

\begin{figure*}
    \includegraphics[scale=0.75]{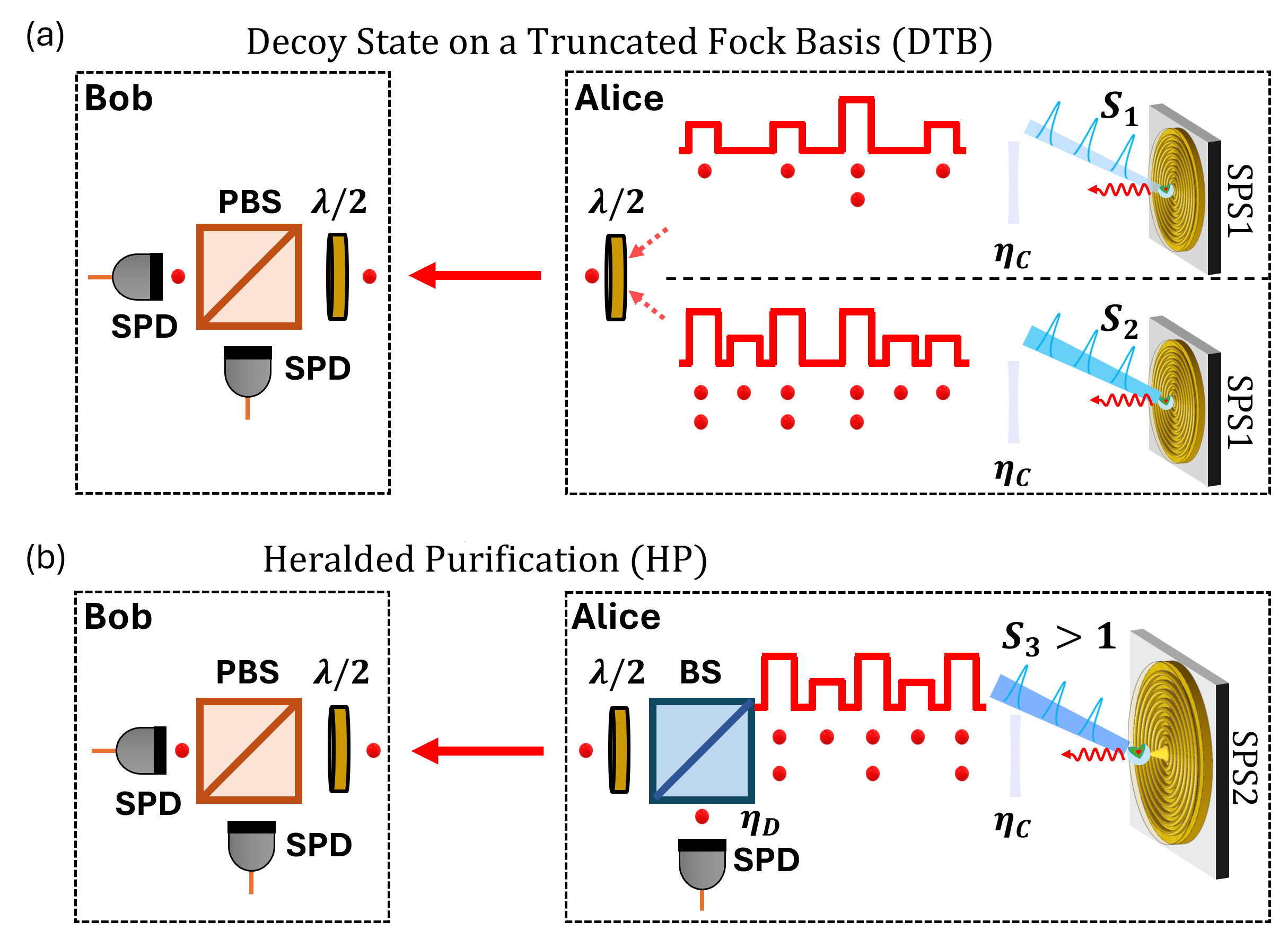}
    \caption{Concept of two protocols based on a high rate, high collection efficiency SPS with imperfect purity:  (a) Experimental setup and sketch of the BB84+DTB protocol, with Alice adjusting the excitation power on the SPS between $S_0=0$, $S_1$ and $S_2$ to control the emitted photon statistics for the 2-decoy protocol, followed by collection optics with efficiency $\eta_C$ and a standard BB84 encoding unit. (b) Experimental setup and sketch of the BB84+HP photon purification protocol. A single excitation power $S_{3}>1$, near the gCQD saturation, is used to excite the SPS followed by collection optics with efficiency $\eta_C$, a beam-splitter (BS), a high efficiency single photon detector with detection efficiency $\eta_D$, and a standard BB84 encoding unit. Here, only same pulse detections at Alice and Bob are used for the secure key. An algorithmic representation of the two protocols is given in Appendix H.}
    \label{fig:fig2}
\end{figure*}

\section{\label{sec:section2}QKD Protocols outperforming WCS}

After establishing the control of the different photon number probabilities, we now utilize this capability for two MCL-enhancing protocols, the first is a decoy-state protocol and the second is a heralded purification protocol.

\subsection{\label{sec:section2a}Decoy State on a Truncated Fock Basis - DTB}

In Fig. \ref{fig:fig2}(a), we present the conceptual experimental setup of the decoy state protocol. Here, an SPS with a near unity photon collection efficiency, $\eta_c\simeq 1$ \cite{Abudayyeh2021SingleNanoantennas} (titled as SPS1), is used for implementing a BB84 protocol, but where Alice controls the excitation powers, $S_0=0$, $S_{1}$ and $S_{2}$. This in turn modifies the photon number statistics, thus allowing for the implementation of decoy and signal states \cite{Lo2005DecoyDistribution,Ma2005PracticalDistribution}.

To establish the DTB protocol, we define the gain of the signal (decoy$_i$) state, $Q_{s}$ ($Q_{d,i}$), as the fraction of encoded photon pulses sent by Alice and detected by Bob (Eq. \ref{eq:gain}). We also define the error rate of the signal (decoy$_i$), $E_{s}$ ($E_{d,i}$), as the fraction of detected encoded photons that had errors (Eq. \ref{eq:error}). The signal pulses correspond to pulses with excitation $S_2$, while the decoy pulses are those excited with $S_1,S_0$. We show in the following that due to the truncated photon number basis of our SPS, only two different decoy states are enough for exact analysis in the framework of the decoy state protocol \cite{Lo2005DecoyDistribution}. 

Within the DTB protocol, Alice can randomly choose to replace the signal pulse with one of the two decoy states with different photon number distributions. These decoy states are not used to create the secure key but to improve the information about the channel and discover any possible eavesdroppers \cite{Ma2005PracticalDistribution}. In the post-processing stage, Alice and Bob can verify the fraction of detected events and the errors for both signal pulses and decoy pulses, thus obtaining the gain and error rates $\{Q_{s},Q_{d,i},E_{s},E_{d,i}\}$. 

On the other hand, due to the uncertainty of the photon-number in each pulse, the gain and error rate for a specific n-photon state, $\{Q_{n},e_{n}\}$, cannot be measured directly and have to be estimated or calculated. Out of this set, the single photon gain ($Q_{1}$) and error rate ($e_{1}$), which are the probability of a detected event (Bob) to originate from a single photon pulse (Alice) and the error rate for single photon pulses respectively are of particular importance. This is since, considering a possible PNS attack by Eve, only information encoded on the single photon pulses is secured. 

Crucially, in our three-intensity DTB protocol, these parameters can be solved exactly since the probability distribution of the signal and decoy states, $\{ P_{n}^{s} \},\{ P_{n}^{d,i}\}$  is known, e.g. Fig. \ref{fig:fig1}(c). To relate the measured values $\{Q_{s},Q_{d,i},E_{s},E_{d,i}\}$  to the unknown set $\{Q_{n},e_{n}\}$, it is useful to define the n-photon yield, $Y_{n}$, as the conditional probability of a detection event given that Alice sent a n-photon state, and therefore $Q_{n}=Y_{n}P_{n}$. With these definitions, the following set of equations can be written \cite{Ma2005PracticalDistribution}: 
\begin{align}\label{calc1}
    Q_{s}&=\sum_{n=0}^{2} P_n^{s}Y_n \qquad E_{s}Q_{s} =\sum_{n=0}^{2} P_n^{s}Y_ne_n \nonumber \\
    Q_{d,i}&=\sum_{n=0}^{2} P_n^{d,i}Y_n \qquad E_{d,i}Q_{d,i} =\sum_{n=0}^{2} P_n^{d,i}Y_ne_n
\end{align}

Importantly, unlike the decoy state protocol implemented for WCS where $n\rightarrow\infty$, which requires an infinite set of decoy states for an exact solution \cite{Lo2005DecoyDistribution}, here $n=0,1,2$ only, due to the truncated sub-Poissonian nature of the gCQD photon emission ($N>2$ photon emissions are neglected). Thus, our equations have only six unknowns: $\{Y_0,\,e_0,\,Y_1,\,e_1,\,Y_2,\,e_2\}$,
suggesting that two decoy states are enough for an exact solution to these equations. This makes the DTB implementation particularly viable, whereas for WCS using two decoy states would only give a bounded approximation for the yield and errors \cite{Ma2005PracticalDistribution}.

The solution of these equations can then be used to estimate the minimum SKR after privacy amplification and error correction, defined as $R_{1}$ for the DTB protocol \cite{Lo2005DecoyDistribution}:
\begin{equation} \label{eq:SKR}
    R_{1} \geq q \{ -Q_{s} f(E_s) H_{2}(E_s)+Q_{1} \left[ 1-H_{2}(e_1)\right]\}
\end{equation}

where $q=0.5$ for the BB84 protocol \cite{Shor2000SimpleProtocol}, $H_{2}$ is the binary Shannon information function \cite{Lo2005DecoyDistribution}  and $f(E_{s})$ is the error correction efficiency (taken to be 1.22 \cite{PhysRevLett.85.1330}).

Assuming $\eta_{C}\simeq1$ for a gCQD coupled to a nanoantenna \cite{Abudayyeh2021SingleNanoantennas} and using the $P_{0},P_{1},P_{2}$ probability values above, we numerically calculate the SKR of such a source  for different channel losses. To do this, we use the channel model connecting the losses, yields, gains and the error rates shown in \cite{Lo2005DecoyDistribution} and in Appendix E with the parameters values (from \cite{gobby2004quantum,Gauthier2013Decoy-stateWaveguide,Abudayyeh2021OvercomingSource,Abudayyeh2021SingleNanoantennas}) presented in Table  \ref{tab:parameters}, Appendix G. Specifically, we set the probability to reach the wrong detector to $e_{d}=3.3\%$ and Bob's detection efficiency to $\eta_{\text{Bob}}=4.5\%$.

\begin{figure*}
    \includegraphics[scale=0.48]{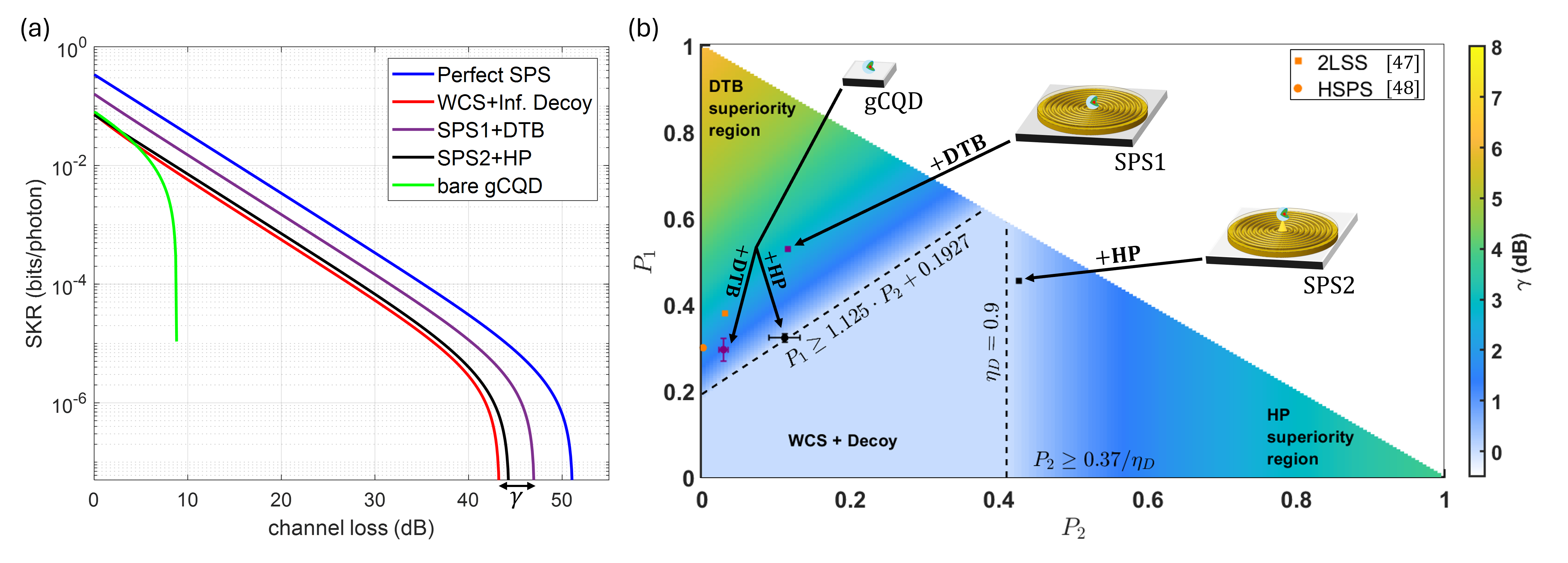}
    \caption{
\textbf{QKD analysis with imperfect single photon sources}. (a) Secure key rate of the BB84 protocol as a function of channel loss for a perfect SPS (blue line), WCS with infinite decoy states and optimized intensities (WCS+Inf. Decoy, red), SPS1 with the decoy state on a truncated Fock basis protocol (SPS1+DTB, purple), SPS2 with the heralded purification protocol (SPS2+HP, black), and a typical gCQD (bare gCQD, green), with a visual representation of $\gamma$, the relative gain. 
The method and parameters of the calculations are found in Table \ref{tab:parameters} in Appendix G, Sec. \ref{sec:section2a}, and Sec. \ref{sec:section2b}. (b) The relative gain of the MCL, $\gamma$, as a function of $P_{1}, P_{2}$ showing two advantageous regimes for SPS with either DTB or HP separated by black dashed lines with their respective conditions. We also present our sources: a bare single gCQD experimentally measured for the DTB protocol (purple circle) and the HP scheme (black circle) at different intensities (hence different photon statistics),  SPS1 in the purification regime (purple square), SPS2 in the decoy state regime (black square), and two, previously analyzed, non-classical sources \cite{Gauthier2013Decoy-stateWaveguide,Wang2008ExperimentalSource} marked with an orange square/dot.
}
    \label{fig:fig5}
\end{figure*}

In Fig. \ref{fig:fig5}(a) we show the calculated SKR under different channel losses for a standard WCS with decoy states in the ideal, simulated, asymptotic case (infinite number of decoy states) with optimized Poissonian distribution parameters, as given by Ref. \cite{Ma2005PracticalDistribution}, setting the detection efficiency and error probability as previously explained. This was compared to an SPS based on a gCQD coupled to a nanoantenna (SPS1) \cite{Abudayyeh2021SingleNanoantennas} using the above DTB protocol with the realistically obtainable probabilities of the signal state for such a device (see Appendix G). A clear improvement in the SKR and the MCL of our imperfect SPS over WCS is seen, with over $3$ dB MCL enhancement, and also an improvement over the best existing cryogenic state-of-the-art SPS \cite{loredo2016scalable,somaschi2016near, he2013demand} or comparable cryogenic, fiber-coupled results \cite{morrison2023single}. We also show a comparison to a perfect single photon source ($P_{1}=1$). Surprisingly, applying our truncated decoy-state protocol yields performance not far worse than a perfect SPS, even though our SPS, operating under ambient conditions, is far from being ideal.

\subsection{\label{sec:section2b}Heralded Purification - HP}
In Fig. \ref{fig:fig2}(b), we present the experimental configuration of the heralded purification protocol (BB84+HP). In this scheme, we consider an SPS consisting of a gCQD on a hybrid nanocone-antenna device (titled here SPS2). Such a device showed both $\eta_{C}\simeq 1$ together with high attainable values of $P_2$, resulting from a large Purcell factor induced by the nanocone  which significantly enhances $QY_{XX}$  \cite{Abudayyeh2021OvercomingSource,Matsuzaki2017StrongAntenna}. Alice operates at a single excitation power $S_{3}>1$, deep in the saturation regime, to excite the BX state with a very high probability, thus maximizing $P_2$, which is now only limited by $QY_{XX}$ and $QY_{X}$ \cite{Abudayyeh2021OvercomingSource}. 
A beam-splitter (BS) and a single photon detector (SPD) are added as a purification stage in Alice's system, in the emission line of the SPS and before the standard BB84 encoding unit. As $P_{N>2}$ are negligible (see Appendix D), a real photon detection event in Alice's detector sets $\tilde{P}_2=0$ for that pulse, where $\tilde{P}_n$ is the effective photon number probability after the purification stage.
In the  sifting step of the QKD protocol, only events with same-pulse detections by Alice and Bob are considered for the secure key, thus eliminating all multiphoton events at a cost of lower signal rates.\newline

\begin{figure*}
    \includegraphics[scale=0.85]{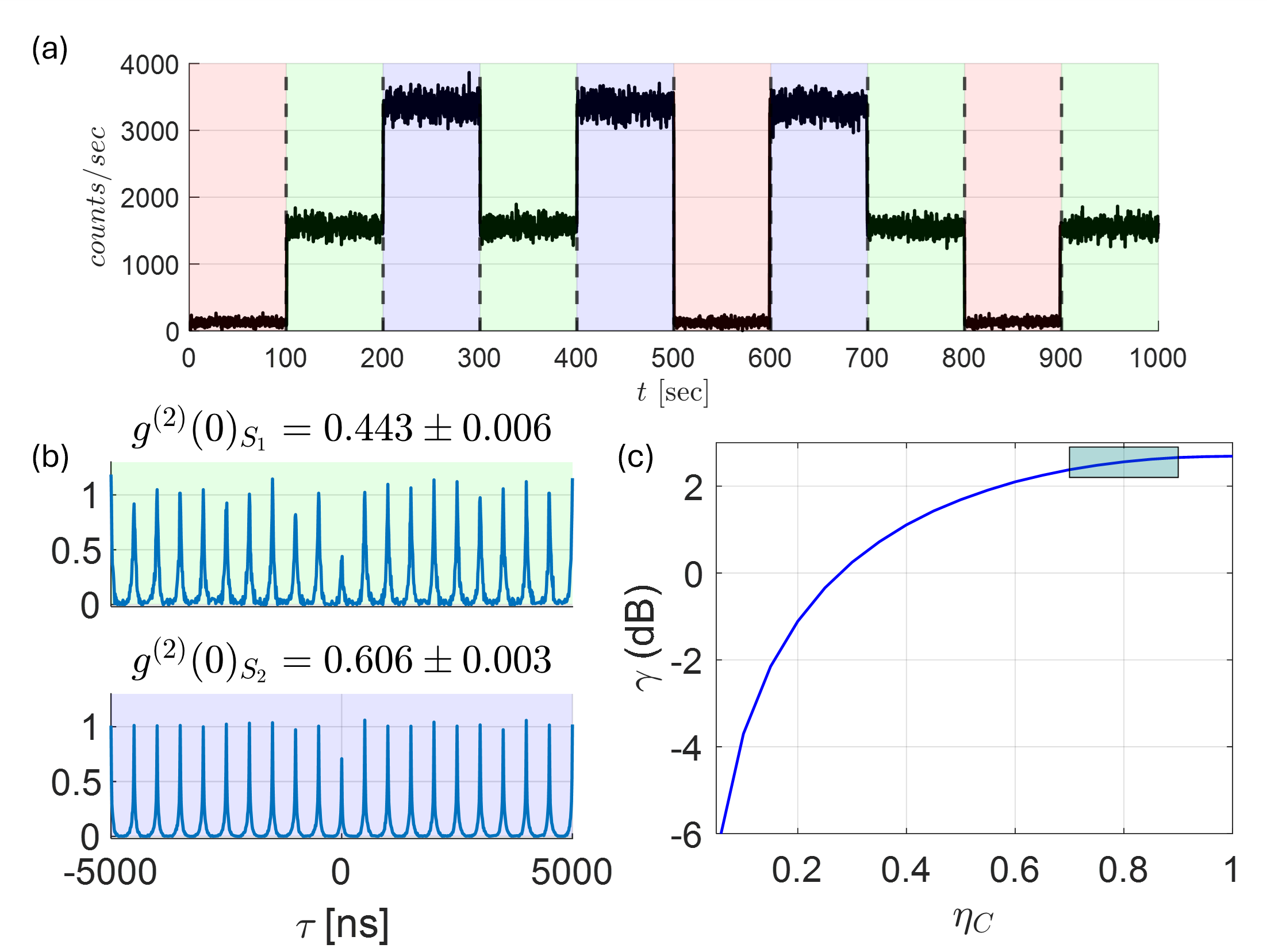}
    \caption{\textbf{Experimental results for emulating the DTB protocol.} (a) Controlling the photon emission counts by the excitation laser intensity (405 nm, 2 MHz rep. rate, 2 ns pulse duration). Red regions mark $S_0$, green regions mark $S_1$ and purple regions mark $S_2$ appearing in  Fig. \ref{fig:fig1}(c). The time steps (100 sec) taken for the power modulation of the excitation laser were chosen to emphasize the feasible control of the photon emission. (b) $g^{(2)}(0)$ for the two different intensities, $S_1$, $S_2$. (c) Calculation of $\gamma$ values as a function of $\eta_{C}$. The blue box represents the $\eta_C$ range demonstrated in our devices \cite{Abudayyeh2021OvercomingSource,Abudayyeh2021SingleNanoantennas}, all with $\gamma>2$ dB.}
    \label{fig:fig3}
\end{figure*}

\begin{figure*}
\includegraphics[scale=0.745]{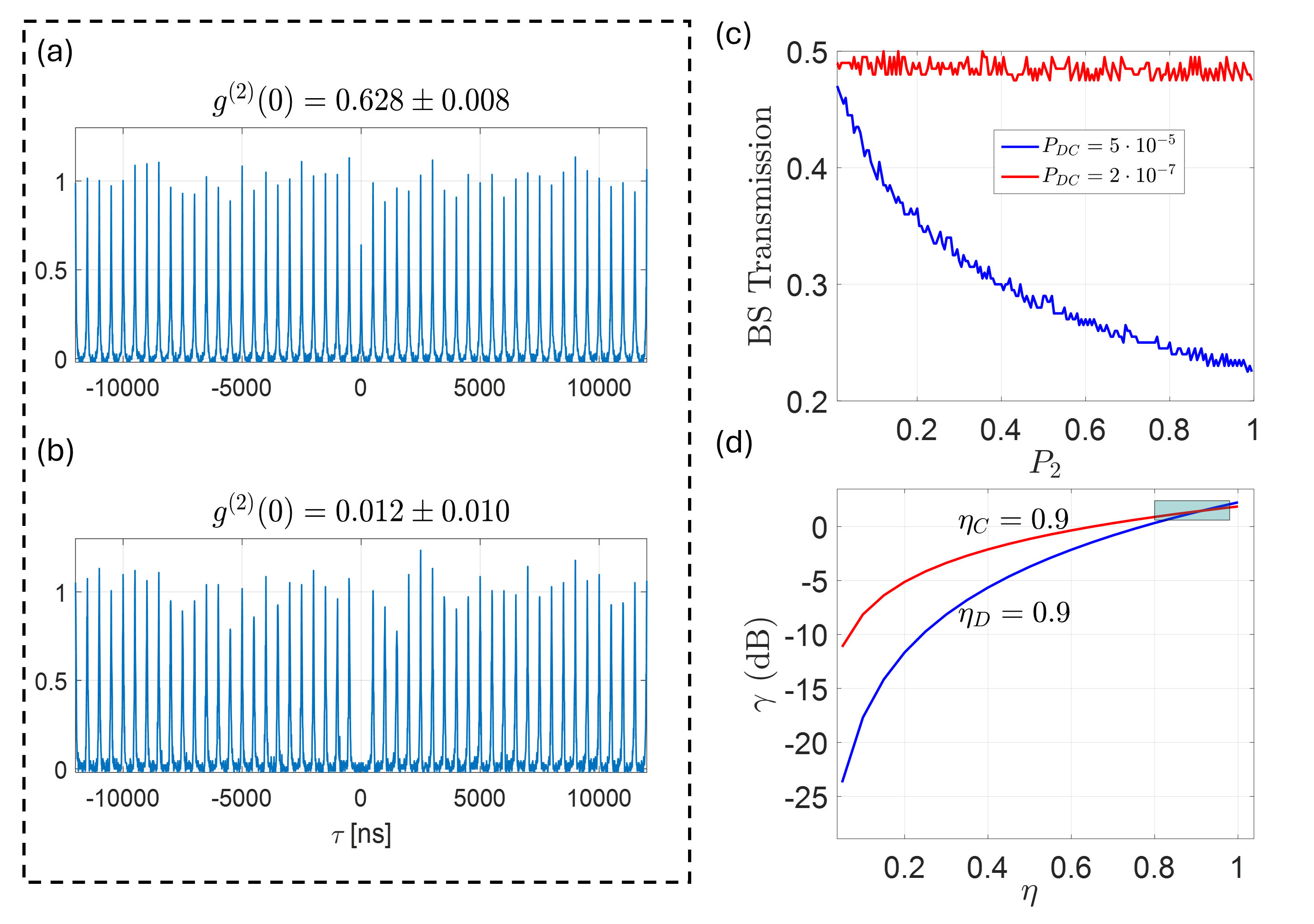}
\caption{\label{fig:fig4} \textbf{Experimental results for emulating the HP protocol.} Second-order correlation measurements of the bare gCQD SPS without (a) and with (b) the HP post-processing scheme, resulting in near-zero $g^{(2)}(0)$. The error is mainly due to the SPD dark noise. (c) Calculation showing the optimal transmission $T$, of Alice's BS, required for maximizing $\gamma$ as a function of $P_2$ of the SPS, for two values of $P_{DC}$, corresponding to 100 dark counts per second of the SPD, and SPS excitation repetition rates of $2$ (blue) and $500$ (red) MHz. (d) Calculation of $\gamma$ as a function of $\eta_{C}$ with a fixed $\eta_D$ (blue line) and as a function of $\eta_D$ with a fixed $\eta_C$ (red line). The blue box represents the range demonstrated in our devices \cite{Abudayyeh2021OvercomingSource,Abudayyeh2021SingleNanoantennas}.}
\end{figure*}

Given an SPS having $\{ P_0,\,P_1,\,P_2 \}$, the effective distribution sent to Bob with the HP protocol depends on the reflectance and transmission ($R$ and $T$) of the BS and the detection efficiency of Alice's detector ($\eta_{D}$), as well as the probability of a dark count at Alice's detector ($P_{DC}$):
\begin{align} \label{eq:HP}
    \tilde{P}_1&=2P_{2}RT(\eta_{D}+P_{DC})+TP_{1}P_{DC} \\
    \tilde{P}_2 &=T^{2}P_{2}P_{DC} 
\end{align}


Given this new distribution, we can again follow the well established method to estimate the SKR of a BB84 protocol (but now implemented with HP) \cite{Lo2005DecoyDistribution,GottesmanSecurityDevices}.
Here, the estimated SKR after privacy amplification and error correction, defined as $R_{2}$ for the HP protocol, is given by \cite{Alleaume2014UsingSurvey}:
\begin{equation} \label{eq:SKR2}
    R_{2} \geq q \cdot Q_{s} \{ -H_{2}(E_s)+\Omega \left[ 1-H_{2}(\frac{E_{s}}{\Omega})\right]\}
\end{equation}
where $\Omega=\frac{\tilde{P}_{1}\cdot Y_{1}}{Q_s}$ is the relative error of the quantum channel, and $H_{2}$, $q$, $E_{s}$, $Q_{s}$ are defined similarly to the DTB protocol. 

The black line in Fig. \ref{fig:fig5}(a) shows a realistic calculation of the SKR using actual parameters of SPS2 measured in Ref. \cite{Abudayyeh2021OvercomingSource} (presented in Table \ref{tab:parameters} in Appendix G). A BS reflectance of $50\%$ was chosen, and we specifically use $P_{DC}=2\cdot {10}^{-7}$ corresponding to 100 dark counts per second for a $500$ MHz signal rate which are both commonly attainable with current technology. Again, as is seen in the figure, the new BB84+HP allows for a higher MCL ($\sim 1$ dB) compared to WCS with infinite decoy, due to the extremely low  $\tilde{P}_2$. Remarkably, this SKR enhancement can be achieved with SPS far from being ideal having a very low single photon purity.

\subsection{\label{sec:section2c}Performance Analysis}

Next, we analyze the expected performance of realistic SPS-based BB84-QKD using the above protocols, in comparison to that of WCS with ideal decoy protocols. In particular, we compare the expected performance of our existing room-temperature high brightness, high collection efficiency SPS \cite{Abudayyeh2021OvercomingSource,Abudayyeh2021SingleNanoantennas}. We define the relative gain in the MCL as $\gamma = (MCL/MCL_{WCS})$, and use this parameter to evaluate the relative performance gain.\newline

Fig. \ref{fig:fig5}(b) shows a colormap of the calculated $\gamma$ for different values of $P_{1},P_{2}$ of the SPS. For high $P_1$ and sufficiently low $P_2$, namely for $P_1>1.125P_2+0.1927$, there is a region (marked as "DTB superiority region") where $\gamma>0 \, \text{dB}$, indicating that the use of an SPS with BB84+DTB protocol is advantageous over WCS with BB84  including infinite decoy states (marked as "WCS+Decoy"). In this region, we highlight in orange dots two known non-classical sources \cite{Gauthier2013Decoy-stateWaveguide,Wang2008ExperimentalSource} along with our device (SPS1) consisting of a gCQD coupled to a metal-dielectric Bragg nanoantenna \cite{Abudayyeh2021SingleNanoantennas} (purple), demonstrating that already existing sources, when combined with the DTB protocol, can outperform even ideal WCS protocols. Notably, the use of DTB allows for an SPS with higher probabilities of two-photon events and shows that bright devices (with fewer vacuum events) can be used for QKD even with single photon purities as low as $\sim65\%$ and $g^{(2)}(0)$ values as high as $\sim0.6$.\newline

On the other hand, for an SPS with high enough values of $P_{2}$, there is a region where BB84+HP is advantageous over WCS+Decoy, as shown in the right corner of Fig. \ref{fig:fig5}(b) (marked as "HP superiority region"), where again $\gamma>0 \, \text{dB}$. 
In the implementation of the HP protocol, most one-photon events are discarded, and the key is composed largely of two-photon emission events, as indicated in Eq. \ref{eq:HP}, where the first term is dominant since $P_{DC} \ll P_{1},\,P_{2}$. Therefore, the HP method is most advantageous in the regime where $P_{2}$ is large, regardless of $P_1$. The negligible probability of two-photon events after purification ($<{10}^{-7}$) and minor contributions of the second and third terms in Eq. \ref{eq:HP} result in an effectively pure source, differing from a perfect SPS only in brightness (through the zero-photon probability).
 
 As indicated by the first term of Eq. \ref{eq:HP}, the probability of sending one photon is linearly dependent on $\eta_{D}$, suggesting that the minimum value of $P_{2}$ required for $\gamma>0 \, \text{dB}$ is inversely proportional to Alice's detection efficiency. Calculations yield this relation, giving the condition $P_{2}\geq 0.37/\eta_{D}$ for a balanced 50:50 BS. The values of $\gamma$ and the separation line between the WCS and HP in Fig. \ref{fig:fig5}(b) are evaluated for the realistic case $\eta_{D}=0.9$, yielding an expected enhancement over WCS when $P_{2}>0.41$. The black dot in  Fig. \ref{fig:fig5}(b) represents an actual SPS device consisting of a gCQD coupled to a Bragg antenna with a plasmonic nanocone (SPS2), demonstrated in Ref. \cite{Abudayyeh2021OvercomingSource}, again showing that existing imperfect SPS+HP can compete with WCS+Decoy.

 Finally, we present experimental results of a bare gCQD sample (shown here in purple and black circles with corresponding error bars), where the photon probabilities were experimentally obtained as explained in Appendix C, and the SKR was extracted separately for each protocol with Eqs. \ref{eq:SKR},\ref{eq:SKR2} respectively. The gCQD, excited at several intensities, exhibits different photon emission statistics for each intensity, thus allowing to demonstrate both the DTB and HP protocols. As seen, even bare gCQDs emitting at room temperature, without any special antennas, can reach a the superior regime over WCS+Decoy.

\section{\label{sec:section3}Experimental Results}

\subsection{Proof-of-Concept Measurements of Protocols}

After theoretically showing that existing SPSs, combined with the two new protocols, can outperform WCS in terms of MCL, here we demonstrate experimental proof-of-concept emulations of both protocols using our bare single gCQD presented in Fig. \ref{fig:fig1}(c) as our imperfect SPS emulator.

To experimentally demonstrate the feasibility of the DTB scheme, we use three different pulsed excitation intensities generated by a 405 nm diode laser, marked as $S_0=0$, $S_1$, and $S_2$ in Fig. \ref{fig:fig1}(c) (see Appendix F for the full experimental information). Fig. \ref{fig:fig3}(a) demonstrates photon emission control showing stable photon counts under each excitation intensity. This allows an easy control of the photon statistics, as shown in Fig. \ref{fig:fig3}(b), which presents the $g^{(2)}(0)$ values for $S_1$ and $S_2$, consistent with the modification of photon statistics shown in Fig. \ref{fig:fig1}, where $P^{(S_{1})}_{1}=0.05$, $P^{(S_{2})}_{1}=0.4$ and $P^{(S_{1})}_{2}=0.0005$, $P^{(S_{2})}_{2}=0.15$.

As shown previously, the gCQD SPS with a near unity collection efficiency can outperform  WCS with decoy states, if implemented with the DTB protocol. In Fig. \ref{fig:fig3}(c), we numerically calculate the expected $\gamma$ of a gCQD SPS device in terms of $\eta_{C}$ (see Appendix I). Remarkably, with the current measured parameters, $\gamma>0 \, \text{dB}$ already for $\eta_C>0.3$, which is is easily attainable even for bare gCQDs, as we show later. Our previously demonstrated gCQD based SPS devices \cite{Abudayyeh2021SingleNanoantennas} has $\eta_C>0.7$ (marked by a blue rectangle) leading to an expected $\gamma>2 \, \text{dB}$, already constituting a significant improvement over WCS+Decoy.

Moving to emulation of the HP protocol, we show experimental results of the purification of the gCQD emission in Fig. \ref{fig:fig4} using a 50:50 BS. The second-order correlation measurements of the gCQD without (Fig. \ref{fig:fig4}(a)) and with (Fig. \ref{fig:fig4}(b)) the HP post-processing protocol are presented. As can be seen, a near-zero $g^{(2)}(0)$, limited only by detector noise is achieved, competing with state-of-the-art demonstrations \cite{Nelson2024ColloidalSources, somaschi2016near}. We note that with HP, the photon rate decreases to $\sim 0.5P_{2}\eta_D$, but the collection efficiency $\eta_{C}$ is not affected.

Using the results in Fig. \ref{fig:fig4}(a)-(b), one can extract the \textit{third-order} correlation measurements ($g^{(3)}(0,0)$) \cite{stevens2014third} to determine $P_{3}$, the probability for a three-photon emission (see Appendix D). Using this method, we find that $P_{3} \sim 10^{-5}$, therefore we conclude that $P_{N>2}$ is negligible compared to $P_1$ and $P_2$, justifying our initial assumptions.

Interestingly, the HP efficiency can be optimized by adjusting $T$, $R$ of Alice's BS, depending on the $P_2$ of the source. In Fig. \ref{fig:fig4}(c) we show a calculation of the optimal BS transmission $T$ to reach the highest MCL, as a function of $P_2$, for two dark-counts probabilities $P_{DC}$. For low $P_{DC}$ the optimal transmittance is roughly $0.5$. However, for higher values of $P_{DC}$, the optimal $T$ is smaller than 0.5 and decreases with increasing two-photon probability, in order to minimize the probability for a dark count at Alices' detector simultaneously with two-photons transmission to Bob. Lastly, in Fig. \ref{fig:fig4}(d) we plot the calculated $\gamma$ for a gCQD-based SPS with the above parameters, as a function of $\eta_D$, for a fixed $\eta_{C}=0.9$, and as a function of $\eta_C$ for a fixed $\eta_D=0.9$. Again, the blue box, representing demonstrated values of SPS devices based on gCQD coupled to nanoantennas, shows an improvement over WCS+Decoy. 

\subsection{\label{sec:bb84meas}Experimental Demonstration of the protocols in BB84-QKD using a gCQD}
\begin{figure*}
\includegraphics[scale=0.65]{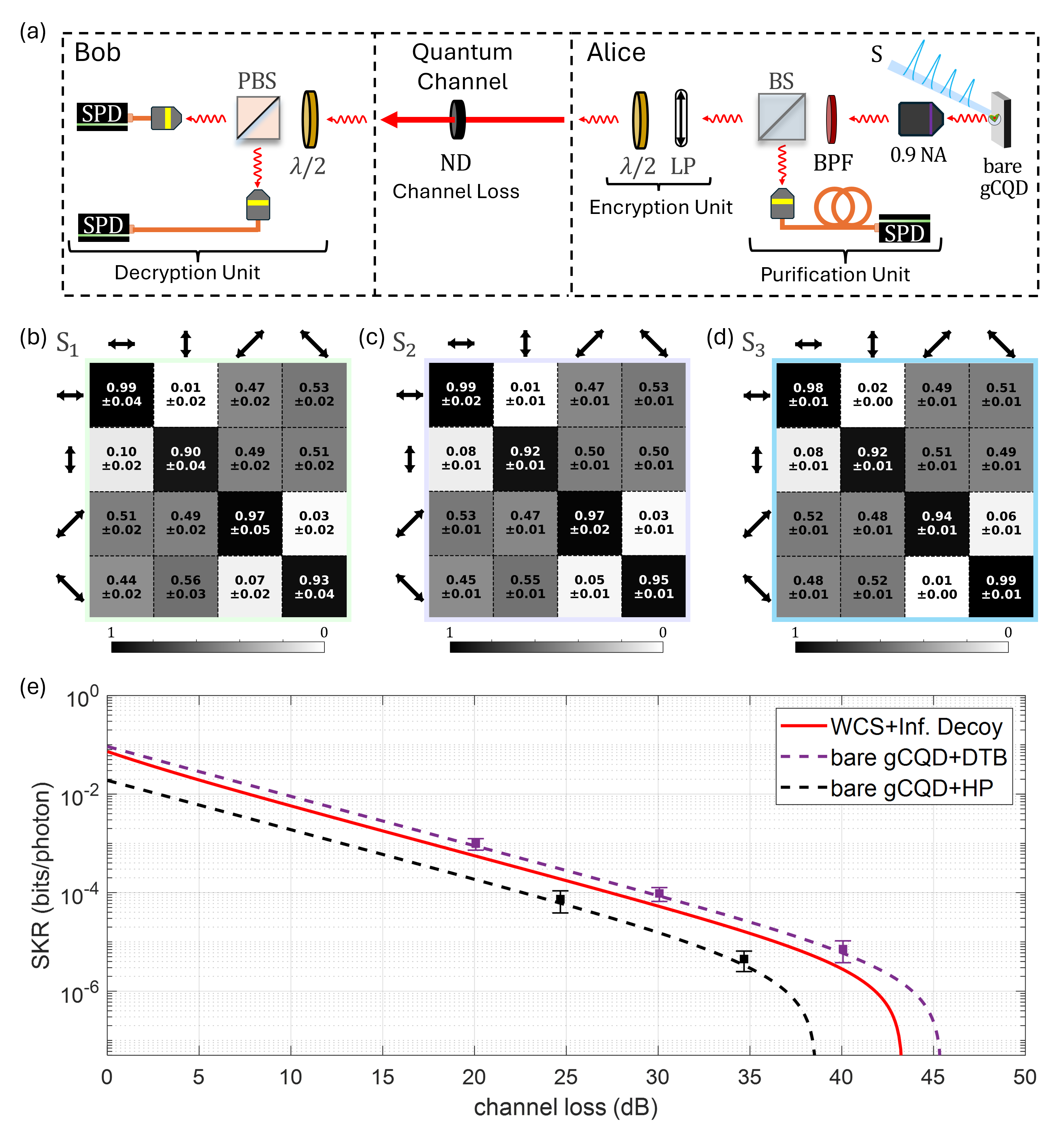}
\caption{\label{fig:fig6} \textbf{Polarization-based BB84 QKD measurements of a bare gCQD sample generated with the two protocols.} (a) The experimental QKD system sketch: Alice's setup includes a purification unit with a beam-splitter (BS) and a single photon detector (SPD), and an encryption unit with a linear polarizer (LP) and half-wave plate ($\lambda /2$). The free-space quantum channel includes ND filters to vary the channel loss. Bob's setup includes the BB84 decryption unit with a $\lambda / 2$, a polarizing beam-splitter (PBS) and two SPDs. Tomography mapping results for different excitation laser intensities: $S_{1}$ (b), $S_{2}$ (c), and $S_{3}$ (d), respectively, corresponding to the decoy (green frame) and signal (purple frame) states in the DTB protocol, and $S>1$ intensity (blue frame) state in the HP protocol. (e) The SKR results extracted from the measurements from a bare gCQD sample  compared to WCS with infinite decoy. The different measured points along  represent the extracted SKR results for several ND filter settings (see full results in Appendix J) using the DTB protocol (purple) and the HP protocol (black). The dashed lines represent the corresponding calculated SKR. The DTB results demonstrate a clear improvement in the maximal channel loss, $\gamma$, over the WCS (red line), while the HP protocol results (dashed black) falls below this due to the small $P_2$ values of bare gCQDs without a nanoantenna.
}
\end{figure*}

Now, we consider a BB84 demonstration using a bare gCQD as the imperfect SPS and utilizing either DTB or HP protocols. 

In Fig. \ref{fig:fig6}(a), we show the QKD system that was used to demonstrate a polarization-based BB84 protocol. The photons emitted from the gCQD sample are collected to Alice's encryption unit, which randomly define both the qubit value and basis \cite{Shor2000SimpleProtocol,bloom2022quantum}. The photons are then propagated through the quantum channel in free-space, where we introduced ND filters in the optical path to simulate channel loss, before arriving at Bob's decryption unit. Here, we define the two mutually unbiased bases as '$+$' ($\ket{H},\ket{V}$) and '$\times$' ($\ket{D}$,$\ket{A}$) \cite{bloom2022quantum}.

By manipulating the different basis settings of both Alice and Bob, the full tomography mapping can be measured. In Fig. \ref{fig:fig6}(b)-(d), we show the tomography results for three different excitation intensities: $S_{1}$, $S_{2}$, and $S_{3}$ without an ND filter in the quantum channel (for the full set of results, see Appendix J).

From these results, the gain $Q_{i}$ and error rates $E_{i}$ for intensity $i$ are extracted using the following relations:

\begin{equation} \label{eq:gain}
    Q_i = \frac{\text{Total \# of detected qubits}}{\text{Total \# of sent qubits}}
\end{equation}

\begin{equation} \label{eq:error}
    E_i = \frac{\text{\# of error qubits}}{\text{Total \# of detected qubits}}
\end{equation}

The total number of detected qubits is the measured counts at both of Bob's detectors at a given basis, and the total number of error qubits is the number of counts measured on the wrong detector (assuming Alice and Bob randomly chose the same basis, according to the BB84 protocol). The total number of qubits sent by Alice is estimated by the following:
\begin{equation}
    \text{Total \# of sent qubits}= N\cdot \eta_{A} \cdot \eta_{C_{NA}}
\end{equation}

where $N$ is the repetition rate of the excitation laser (2 MHz, see Appendix F), $\eta_A\simeq 19.5\%$ is the transmission of Alice's setup measured in our experiments, and $\eta_{C_{NA}}\simeq14.18\%$ is the transmission of the photon collection optics from the gCQD to Alice extracted previously in Ref. \cite{Bloom2024RadialPol}.

The $Q_{i}$, $E_{i}$, along with the measured photon statistics $\{P_0,P_1,P_2\}$ for each intensity $i$, were inserted into Eq. \ref{calc1} to extract $Y_n$ and $e_n$, which were then used for calculating the SKR in Eqs. \ref{eq:SKR},\ref{eq:SKR2} for the DTB and HP protocols respectively. The DTB model requires two decoys ($S_0$ and $S_1$), where $S_0$ is the vacuum state with known parameters (see Appendix G). Here, $S_{2}$ is used as the signal state for the DTB protocol, and $S_{3}$ is used for the HP protocol. 

In Fig. \ref{fig:fig6}(c), we plot the SKR of the gCQD measured with the DTB protocol and the HP protocol, for several channel losses set with various ND filters, compared to WCS with infinite decoy states. A clear improvement of $\sim 2$ dB is demonstrated with the DTB protocol, which agree well with the theoretical curve. As expected, the results with the HP protocol do not show an improvement compared to WCS with decoy, due to the low $P_{2}$ in bare gCQD as compared to those obtained in a full SPS2 devices \cite{Abudayyeh2021OvercomingSource}, nonetheless, showing the feasibility of the two enhanced-QKD protocols with already available SPSs, and their advantage over the best existing WCS solutions.

\section{\label{sec:summary}Conclusions}

As an alternative to the very challenging push for nearly ideal SPS that is required to outperform the existing WCS protocols for QKD, we proposed and experimentally analyzed two simple-to-implement protocols that can allow for even far from ideal SPS, which are currently technologically ready, to beat the state-of-the-art performance of weak coherent states with decoy protocols, achieving over $>3$ dB enhancement in terms of the secure key rate.

The protocols are based on the simple ability to control the statistical distribution of the truncated photon number basis $\{\ket{0},\ket{1},\ket{2}\}$ of a QD BX-X cascaded emission, by varying the excitation power. We showed that depending on the possible attainable values of  $P_1$ and $P_2$, either a decoy state protocol, DTB, or an heralded purification protocol, HP, can be employed to a non-ideal SPS. 

This is a particularly attractive route for improving the performance of current QKD systems, as we have shown that even room temperature, on-chip, compact, and easily integrated SPS devices, such as those based on gCQD coupled to nanoantennas, are already well within the parameter range for superior performance over WCS with decoy states by employing either a DTB protocol or HP. Both protocols have very simple requirements and their application is very general, thus we believe they can be employed efficiently on a vast range of sub-Poisson, quantum emitters, opening a practical and realistic way to implement novel photon sources with superior QKD performance, without the stringent requirements that hindered their practical integration into real-world QKD systems.

Our protocols, which improves QKD performance, could also enhance other quantum cryptography technologies. By improving eavesdropping detection, it could strengthen quantum secure direct communication \cite{hu2016experimental}, quantum secret sharing \cite{hillery1999quantum}, and quantum secure computation \cite{sulimany2024quantum}. These protocols offer versatile improvement across various quantum cryptographic applications.

\section*{Acknowledgments}
The gCQD synthesis was conducted at the Center for Integrated Nanotechnologies (CINT), a Nanoscale Science Research Center and User Facility operated for the U.S. Department of Energy (DOE), Office of Science (SC), Office of Basic Energy Sciences (BES). R.R., Y.B, Y.O., T.L., and K.S. acknowledge the financial support from the Quantum Communication consortium of the Israeli Innovation Authority. J.A.H. was supported in part by the U.S. Department of Energy (USDOE), Office of Science (OS) Office of Advanced Scientific Computing Research, through the Quantum Internet to Accelerate Scientific Discovery Program, and E.G.B. was supported by the USDOE-OS Basic Energy Sciences, through the Center for Integrated Nanotechnologies.

\section*{Author Contributions}
R.R. conceptualized the protocols and  supervised the project. Y.B. and T.L. conducted the experiments. Y.B. fabricated the samples. Y.O. provided the theoretical and numerical analysis. J.A.H. developed the gCQDs and supervised the program at CINT. E.G.B. performed gCQD synthesis and characterization. K.S. assisted the theoretical analysis. All authors contributed to the writeup of the manuscript.

\appendix
\section{Materials and Methods}
Giant colloidal nanocrystal quantum dots (gCQD) of CdSe/CdS core-shell type were used as the quantum emitters in this work. The gCQD core has a diameter of $\sim 3$ nm, while the shell has a diameter of $\sim 15$ nm. The emission wavelength of the gCQD is centered around $650$ nm at room temperature. The properties of these quantum dots was investigated in many works \cite{Colloidal_QDS, Chen2008GiantBlinking, Htoon2010HighlyNanocrystals, Garcia-Santamaria2009SuppressedPerformance}.

\begin{figure*}[!thb]
    \includegraphics[scale=0.65]{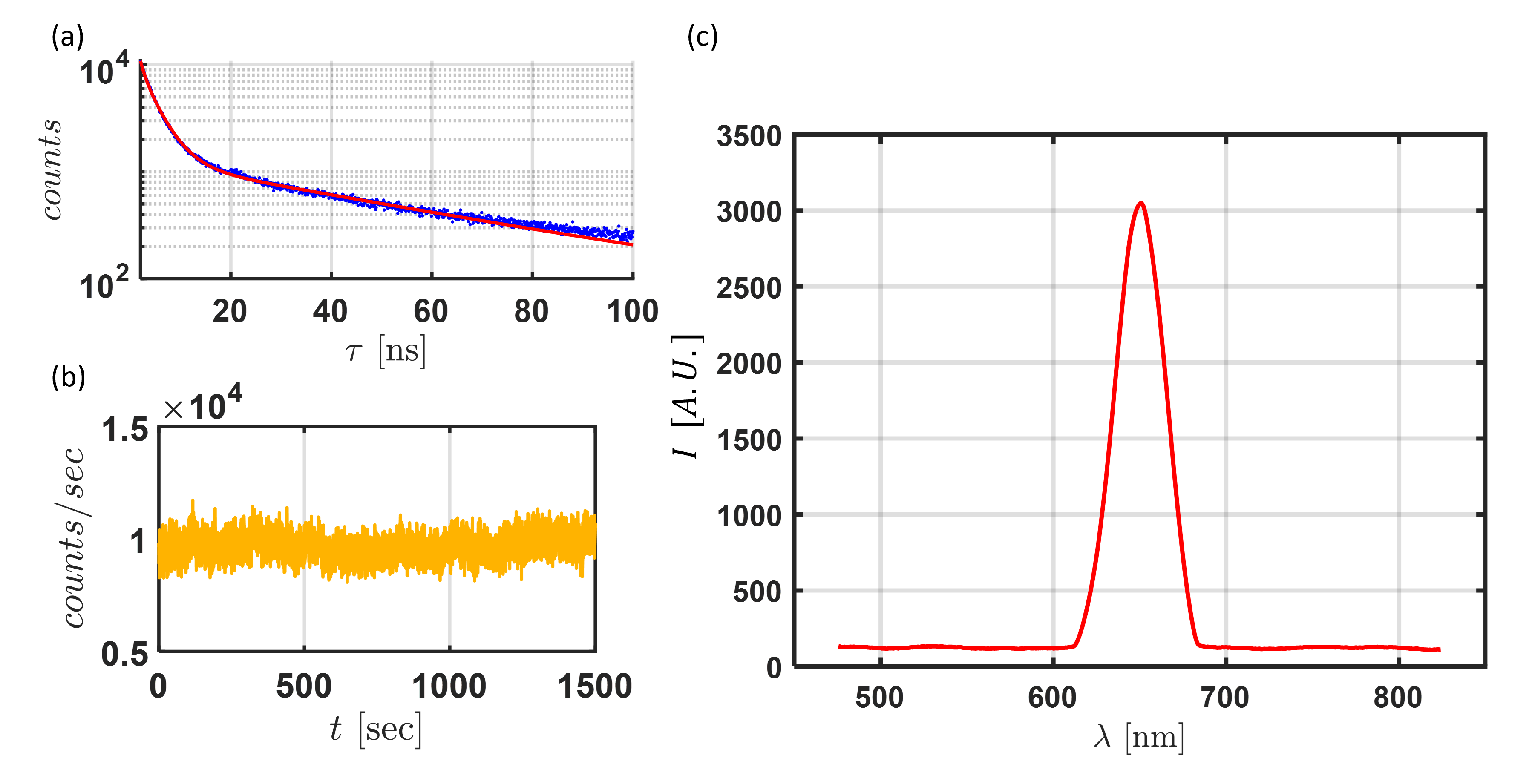}
    \caption{Bare single gCQD properties. (a) Lifetime measurement of a single gCQD on glass, with extracted values of $\tau_{X}=49.3 \pm 3.7$ [ns] and $\tau_{XX}=3.58 \pm 0.16$ [ns], for the X and BX emission respectively. (b) Stability measurement of the photon emission from the gCQD, demonstrating highly stable and non-blinking emission. (c) Spectrum of the gCQD corresponding to the emission both from the X and BX states, centered around $\sim 650$ nm.}
    \label{fig:supp2}
\end{figure*}

In Fig. \ref{fig:supp2}, we demonstrate several properties of single gCQDs. The extracted lifetime of the gCQD from a bi-exponential fit due to the emission both from the BX and the X states in presented in Fig. \ref{fig:supp2}(a), demonstrating ns radiative transitions on bare gCQDs. As investigated in \cite{Bloom2024RadialPol, Abudayyeh2021OvercomingSource}, the plasmonic coupling of the gCQD to the metallic resonator shortens the photon lifetime to $\sim 10-100$ ps. The stability of the bare gCQD is shown in Fig. \ref{fig:supp2}(b), with a stable emission of a single gCQD on glass at room temperature, exhibiting a non-blinking emission for long times \cite{Chen2008GiantBlinking}. In Fig. \ref{fig:supp2}(c), we plot the spectrum of the gCQD, with a broad emission (FWHM of $\sim 25$ nm) at room temperature. This demonstrates the spectral overlapping between the BX and X states at high temperatures.

The gCQDs were initially diluted in a Hexane and polymethyl methacrylate 495 A5 (PMMA) solution with sparse ratios among each material (1:200:5000), to ensure the distribution of single gCQDs on the sample. To achieve this, we used an iterative method where different ratios were tested, stirred with a shaker and then spin-coated on a glass slide using a two-step process (500 RPM for 5 seconds, 4000 RPM for 40 seconds),

The plasmonic device was fabricated using methods similar to those presented in \cite{Bloom2024RadialPol, Abudayyeh2021OvercomingSource, Abudayyeh2021SingleNanoantennas}, employing the template stripping method \cite{stripping} to create an Au metallic device consisting of a bullseye concentric antenna with and without a nanocone resonator. The gCQDs were coupled to the nanoantennas using fabrication methods similar to those described in \cite{qd1, qd2, Abudayyeh2021SingleNanoantennas, Bloom2024RadialPol}.

\section{Probability Distribution and Quantum Yields}
In Sec. II of the manuscript, we present a model for the probabilities to excite one or two excitons depending on the excitation laser power. One can use this model to combine the emission distribution and the quantum yields (QYs) of the source. 

Given that a single exciton was excited, the probability to emit two photons is zero, while the probability to emit one photon is given by the exciton quantum yield, $QY_{X}$. Given that two excitons are excited (as in the biexciton state), the probability to emit two photons is the probability that both the exciton and the biexciton recombined radiatively, suggesting that the probability is $QY_{X}QY_{XX}$. Therefore, the probability of a single photon emission is the probability that either the exciton or the biexciton recombined radiatively, giving $QY_{X}+QY_{XX}-2QY_{X}QY_{XX}$.

The complete photon emission distribution is given by:
\begin{align}
    P_{2} &= P_{XX}(I)QY_{XX}QY_{X} \\
    P_{1} &= P_{XX}(I)(QY_{X}+QY_{XX}-2QY_{X}QY_{XX})+ \\&+P_{X}(I)QY_{X}  \nonumber \\
    P_{0} &= 1-P_{1}-P_{2} \label{p0new}
\end{align}

A quantitative metric for the goodness of the fit, the normalized root mean square error \cite{shcherbakov2013survey} (NRMSE), was generated by comparing the theoretical model for the probabilities with the saturable behavior of the source. In our case, the fit gave a result of of NRMSE $\leq 0.012$, showing a good correlation between the theoretical model and experimental results.

\section{Probability extraction from a truncated Fock basis} \label{Supp:one}

The second order correlation function is given by:
\begin{equation*}
    g^{(2)}(\tau)=\frac{\left< a^{\dagger}(t) a^{\dagger}(t+\tau) a(t+\tau) a(t)\right>}{\left< a^{\dagger}(t)a(t) \right> \left< a^{\dagger}(t+\tau) a(t+\tau) \right>}
\end{equation*}
where $a,a^{\dagger}$ are the annihilation and creation operators.
For a stationary source $\left( \left< n(t) \right>=\left< n(t+\tau) \right> \right)$ \cite{fox2006quantum}, the zero delay correlation function can be written as:
\begin{align} \label{eq:g20}
    g^{(2)}(0) & =\frac{\left< n(n-1) \right>}{\left< n \right> ^{2}}= 
    \frac{\left< n^{2}-n \right>}{\left< n \right> ^2}=\frac{\left< n^{2}\right> - \left<n \right>}{\left< n \right> ^2} \\
    & = \frac{\left< n^{2}\right> +\left< n \right>^2 - \left< n \right> ^2- \left<n \right>}{\left< n \right> ^2} = 1+ \frac{Var(n)-\left< n \right>}{\left< n \right> ^2} \nonumber
\end{align}
In this way, we describe the second order correlation function at zero delay as a relation between the distribution's mean and variance, where $\left< n \right> =P_{1}+2P_{2}$, $\left< n^2 \right>= P_{1}+4 P_{2}$ and $Var(n) =\left< n^2 \right>-\left< n \right>^2 $.\\

With the transmission of the system ($\eta$) and the laser's repetition rate ($N$), the photon detection rate, $C$, is given by:
\begin{equation} \label{eq:countrate}
    \frac{C}{N}=\sum_{n=0}^{\infty} \eta_{n} P_{n} \approx \sum_{n=0}^{\infty}  \eta  n P_{n} = \eta \left< n\right>
\end{equation}
where $\eta_{n}=1-(1-\eta)^{n}$ is the detection event probability of the n-photon state \cite{Lo2005DecoyDistribution}, which can be approximated in the limit of $\eta\ll 1$ ($\sim 10^{-2}$ in our case) to $\eta_{n}\approx n \eta$.

This allows to experimentally determine $P_{0}$ using the following:
\begin{equation}\label{eq:p0_eqapp}
    P_{0} = 1 - \frac{C}{\eta \cdot N}
\end{equation}

Using Eqs. \ref{eq:g20},\ref{eq:p0_eqapp} we can define two equations that relate $P_{0}$ and $g^{(2)}(0)$ to $P_{1}$ and $P_{2}$:
\begin{align} \label{eq:findingprob}
        P_{0} + &P_{1} + P_{2} = 1 \\
        g^{(2)}(0) &= \frac{2P_{2}}{(P_{1} + 2P_{2})^{2}}
\end{align}

The experimentally measured $P_0$ and $g^{(2)}(0)$ are used to solve these equations for $P_1$ and $P_2$, and to obtain the first two moments of the distribution. In our case, this contains all the required statistical information, thus allows us to extract the whole probability distribution for the truncated Fock basis from just two measured quantities. \\

\section{$P_{N\geq3}$ probability extraction from $g^{(3)}(\tau_{1},\tau_{2})$ measurements}

By introducing another detector to our system, as explained for the HP protocol (Sec. IIIB), we can use a similar set of equations with minor adjustments (as in Appendix C) to estimate $P_{3}$, the probability of an emission of three photons from the source. This allows us to estimate the probability of higher photon probabilities, to justify our assumptions regarding the photon statistics. 

In this case, the third-order correlation function at zero delay can also be described by the relation between the distribution mean and variance. Here, $\left< n \right> =P_{1}+2P_{2} + 3P_{3}$, $\left< n^2 \right >= P_{1}+4P_{2} + 9P_{3}$, and $\left< n^3 \right>= P_{1}+8P_{2} + 27P_{3}$. In addition, for a stationary source, $g^{(3)}(0,0)$ is given by \cite{stevens2014third}:

\begin{equation}\label{g3}
    g^{(3)}(0,0) = \frac{\braket{n(n-1)(n-2)}}{\braket{n}^{3}} = \frac{\braket{n^3}-3\braket{n^2}+2\braket{n}}{\braket{n^3}}
\end{equation}

Now, by using Eqs. \ref{eq:g20},\ref{p0new},\ref{g3} we have another set of equations that relate $P_{0}$, $g^{(2)}(0)$ and $g^{(3)}(0,0)$ to $P_{1}$, $P_{2}$ and $P_{3}$:
\begin{align} \label{eq:findingprob2}
        P_{0} + &P_{1} + P_{2} + P_{3} = 1 \\
        g^{(2)}(0) &= \frac{2P_{2} + 6P_{3}}{(P_{1} + 2P_{2}+3P_{3})^{2}} \\
        g^{(3)}(0,0) &= \frac{6P_{3}}{(P_{1} + 2P_{2} + 3P_{3})^{3}}
\end{align}

Where we assumed $P_{N>3}=0$.

Therefore, one can extract $P_{1}$, $P_{2}$ and $P_{3}$ from the measured $P_{0}$, $g^{(2)}(0)$, and $g^{(3)}(0,0)$, which gives the first three moments of the distribution. In our case, the third-order correlation at zero delay was measured in the $S>1$ regime, with the highest probability for higher multi-exciton emission. The measured result of correlations between the three detectors was found to be: 
\begin{equation}
    \boxed{g^{(3)}(0,0)=0.00065}
\end{equation}

With a measured $g^{(2)}(0) = 0.747\pm0.003$ and $P_{0}=0.57\pm0.030$. Plugging all these into Eq. \ref{eq:findingprob2} gives:
\begin{align} \label{eq:findingprob3}
        P_{1} = 0.3233 \pm 0.0094 \\
        P_{2} = 0.1112 \pm 0.0207 \\
        P_{3} = 0.0000177 \pm 0.00000052
\end{align}

Therefore, we conclude that $P_{3} \ll P_{1},P_{2}$, leading to the conclusion that for $N\geq3$ photons the probability is negligible. This finalizes our assumptions regarding the characterization of the photon statistics.

\section{DTB Model}
In this section, we present the model used for the estimation of the SKR for the decoy state protocol on a truncated basis (DTB) that was presented in the paper. With the probability distributions for the signal and the decoy states, $\{ P_{n}^{s} \}_{n=0}^{2}, \,\, \{ P_{n}^{d,i} \}_{n=0}^{2}$, and the n-photon yield, $Y_{n}$, the gains are given by:
\begin{align}
    Q_{s} &= Y_{0}P_{0}^{s}+Y_{1}P_{1}^{s}+Y_{2}P_{2}^{s} \\
    Q_{d,i} &= Y_{0}P_{0}^{d,i}+Y_{1}P_{1}^{d,i}+Y_{2}P_{2}^{d,i}  
\end{align}

Where $i$ represents the two corresponding decoy states.\newline

Similarly, the error rates are:
\begin{align}
    E_{s}Q_{s} &= Y_{0}P_{0}^{s}e_{0}+Y_{1}P_{1}^{s}e_{1}+Y_{2}P_{2}^{s}e_{2} \\
    E_{d,i}Q_{d,i} &= Y_{0}P_{0}^{d,i}e_{0}+Y_{1}P_{1}^{d,i}e_{1}+Y_{2}P_{2}^{d,i}e_{2}  
\end{align}
where $e_{n}$ is the n-photon error rate.
According to the channel model, the n-photon yield is given by the n-photon transmission of the channel, $\eta_{n}$, and the dark-count probability, $P_{DC}$:
\begin{equation}
    Y_{n}= \eta_{n}+P_{DC}- \eta_{n}P_{DC}
\end{equation}
where $\eta_{n}$ is given by $\eta_{n}= 1-(1-\eta)^{n}$ such that $\eta$ is the overall channel transmittance.
Finally, the n-photon error rate expression for our model is:
\begin{equation}
    e_{n}= (e_{d}\eta_{n}+\frac{1}{2}P_{DC})/Y_{n}
\end{equation}
where $e_{d}$ is the probability to reach the wrong detector. Using $Q_{1}=Y_{1}P_{1}$, where $Y_{1}$ can be experimentally calculated by solving Eqs. E1-E4, as explained in Sec. IV.B. In the numerical performance, presented in Fig. \ref{fig:fig5}(a), the gain is estimated using Eq. E5, by measuring the dark count probability $P_{DC}$ in our system and setting the overall transmission $\eta$. This gives all the required parameters for $R_{1}$ (Eq. \ref{eq:SKR}), which are valued and can be inserted to yield the bound of the SKR. To obtain the secure key rate per channel loss, we iterate through different transmissions.

\section{Experimental Details}
\begin{figure*}[!thb]
\includegraphics[scale=0.58]{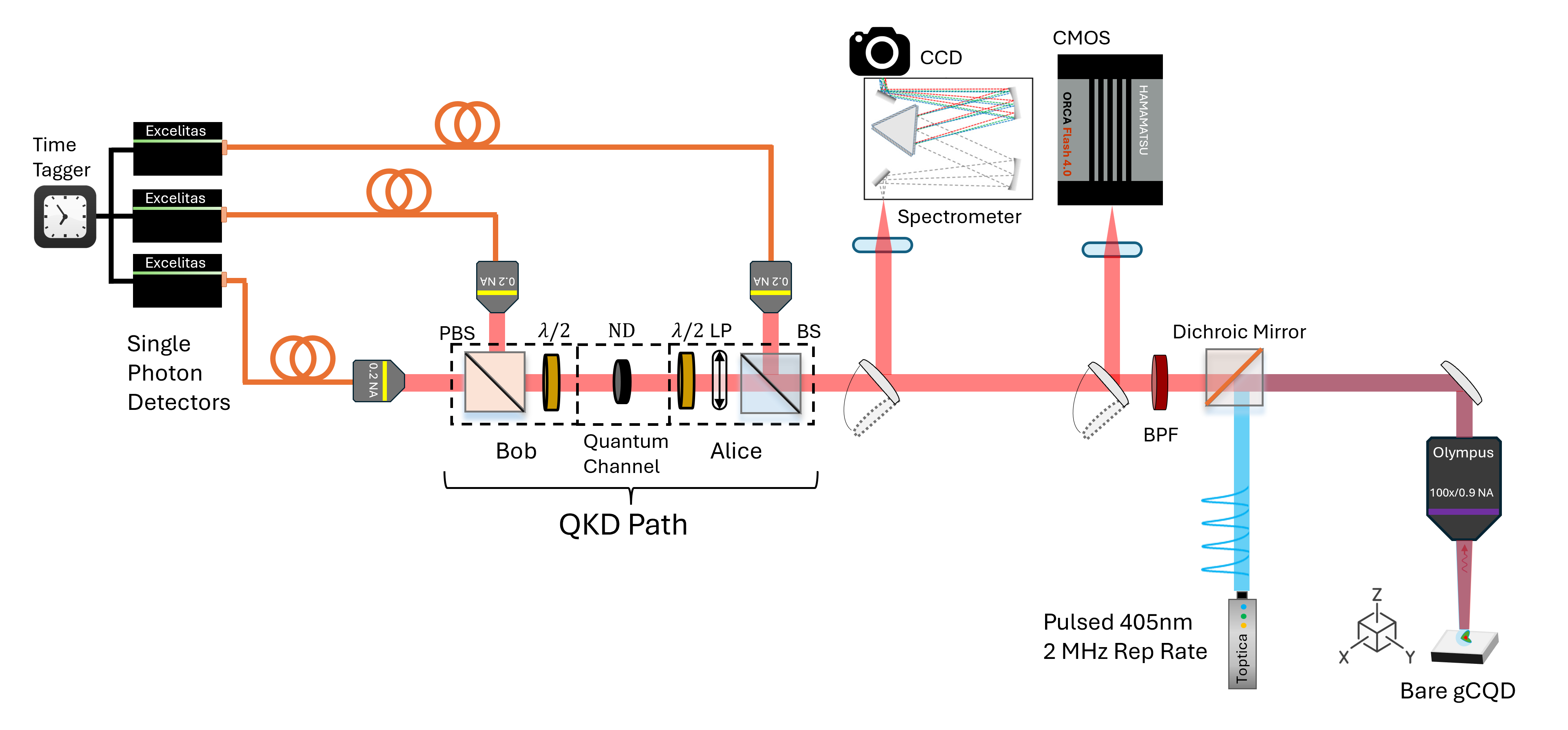}
\caption{\label{fig:supp1} \textbf{Experimental setup.} Excitation with a 405 nm diode pulsed laser operating at 2 MHz was focused on the gCQD based sample, emitting photons to the channel. Correlation measurements, stability and saturation curves were measured using the single photon detectors after the HBT setup. The CMOS camera and spectrometer were used to characterize the gCQD emission. The DTB protocol (shown in Fig. 2(a)) was demonstrated using different intensity powers for the excitation laser. The HP scheme (shown in Fig. 2(b)) was demonstrated using a third detector introduced in the channel along with a 50:50 BS. A standard BB84 encryption (Alice) and decryption (Bob) unit was also introduced in the system.}
\end{figure*}
The optical setup for correlation measurements and the BB84 QKD protocol demonstration is illustrated in Fig. \ref{fig:supp1}. A diode laser (Toptica IBeam Smart) operating at a wavelength of 405 nm generated pulses at a repetition rate of 2 MHz. The excitation laser properties were chosen to allow for a full relaxation of the biexciton and exciton states through the radiative transition channels \cite{orfield2018photophysics, singh2023inside}, thus eliminating unwanted non-radiative effects.

In the DTB scheme, the excitation power was tuned between three different intensities, one which is zero, to obtain information on the photon statistics for each pump. The two other intensities correspond to an average power of $S_{1} \sim 0.01$ mW and $S_{2} \sim 0.05$ mW.

In the HP scheme, the excitation power was set to a high intensity as described in the manuscript, corresponding to an average power of $S_{3} \sim 0.15$ mW.

The excitation laser was focused on the gCQD sample with a 0.9 NA objective (Olympus MPLFLN100xBD), and was scanned using Galil and Zaber electrical stages. The emission from the gCQD devices was collected using the same objective and spectrally filtered from the excitation laser using a 567 nm long-pass dichroic mirror.

Photoluminescence (PL) measurements were performed using a Hamamatsu CMOS camera to identify single gCQDs and the SPS devices. A white light source was introduced to the optical setup alongside the excitation laser path, to scan the device area and located PL from active emitters.

To ensure the emission originated from gCQD based devices, the emission was directed to a spectrometer (Princeton SpectraPro 2500) connected to a CCD camera (PIXIS 256BR), verifying that the emission is centered around the gCQD emission energy \cite{Abudayyeh2021OvercomingSource, Bloom2024RadialPol}.

For time-resolved single photon correlation measurements, the emission was directed to a Hanbury-Brown Twiss (HBT) \cite{HBT} module, consisting of a beam-splitter (BS) and a set of single photon detectors (Excelitas SPCM-AQRH-14-FC), referred to as Bob's detectors, which were coupled to the system using multimode fibers. The signal from each detector was routed to different channels in the time tagging instrument (Swabian TimeTagger 20). The time tagger provided output of all arrival times and channel labels of photons within the set of exposure times, commonly referred to as global times.

Another channel in the time tagger recorded the excitation pulse times, commonly referred to as local times, which served as a trigger channel for the detectors. The histogram of local times is required for lifetime extraction \cite{Abudayyeh2021OvercomingSource}.

Specifically in the HP scheme, a second BS was introduced to the system and coupled to a third single photon detector (referred as Alice's detector). During a post-processing step, data was retained only when both Alice's detector and either of Bob's detectors registered a photon appearance in the same pulse. This was determined using the local times issued for each channel in the time tagger recorded data. The dark count rate of the single photon detectors, as provided by the manufacturer and verified experimentally, is approximately 100 counts per second.

\section{Parameter Table for the DTB and HP Protocols Performance Analysis}
\begin{table}[ht]
\centering
\normalsize
\begin{tabular}{|c|c||c|c|}
\hline
 Source & Protocol&$P_{1}$ & $P_{2}$    \\ \hline
SPS1  & DTB& $0.529$   & $0.112$   \\ \hline
SPS2  &HP& $0.458$   & $0.427$    \\ \hline

bare gCQD  &DTB - decoy ($S_{1}$)& $0.096$   & $0.0017$   \\ \hline

bare gCQD  &DTB - signal ($S_{2}$)& $0.296$   & $0.029$     \\ \hline

bare gCQD  &HP ($S_{3}$)& $0.3231$   & $0.1114$   \\ \hline
\end{tabular}
\caption{Realistic parameter values for the analysis presented in Fig. \ref{fig:fig5}(a)-(b) and Fig. \ref{fig:fig6}(c), done for the DTB protocol using a gCQD-based nanoantenna device \cite{Abudayyeh2021SingleNanoantennas} (SPS1), the HP protocol using a gCQD-based nanocone and nanoantenna device \cite{Abudayyeh2021OvercomingSource} (SPS2), and an experimental demonstration of both protocols using a bare gCQD.  $P_{1}$, $P_{2}$ denote the one- and two- photon probabilities respectively.}
\label{tab:parameters}
\end{table}

In the experimental and numerical analysis, the realistic dark count probability is given as $P_{DC}=2\cdot10^{-7}$ as explained in the manuscript. In addition, $\eta_{C}=1$ and $\eta_{D}=0.9$ define the optimal but realistic collection and detector efficiencies, as explained previously. Lastly, $Y_{0}=1.7\cdot10^{-6}$, $e_{0}=0.5$ denote the vacuum yield and error rate respectively, are used for the vacuum decoy signal in the DTB protocol, $S_{0}$. In addition, the probability to reach the wrong detector was set to $e_{d}=3.3\%$ and Bob's detection efficiency to $\eta_{\text{Bob}}=4.5\%$, as explained in the main text. The $P_{1}$ and $P_{2}$ parameters for different sources are presented in Table \ref{tab:parameters}.

The error bars for the photon emission probabilities ${P_1 , P_{2}}$ were calculated by error propagation accounting for the $g^{(2)}(0)$ and $P_{0}$ deviations. In addition, the error bars for the SKR were extracted by error propagation, considering the gain and error rate uncertainties as extracted from the tomography maps. These uncertainties were measured by the deviation from a mean count value over long exposure times, for each row in the tomography map.

\section{Algorithmic Representation of the QKD protocols}
The algorithm representation of the DTB protocol and the HP protocol is given in Algorithm 1 and Algorithm 2 respectively. 

\begin{algorithm}[H]
\caption{Decoy State on a Truncated Fock Basis Protocol (DTB)}\label{alg:alg1}
\begin{algorithmic}[1]
\Require {Realistic quantum emitter with photon statistics $\{P_0, P_1, P_2\}$, controlled laser excitation powers of $S_0$, $S_1$, $S_2$}
\Ensure Ensure a shared secure key between Alice and Bob using decoy states
\Statex \textbf{Initialization:} Set laser excitation powers to $S_0$, $S_1$, $S_2$; according to the photon statistics at each power.
\For{each clock cycle defined by the excitation laser's pulse repetition rate}
    \State \textbf{Alice randomly selects an excitation power}:
    \Statex \quad {$S_2$ as the signal state with a single photon probability of $P^{(s)}_{1}$}
    \Statex \quad {$S_1$ as the first decoy state}
    \Statex \quad {$S_0$ as the second (vacuum) decoy state}

    \State Alice excites the device with the selected excitation power. The photons emitted from the device are propagated in free-space to Alice's encryption unit.
    \State Alice encodes the photon using the BB84 encryption unit and sends the photon to Bob through the free-space quantum channel.
\EndFor
\State Bob measures each received photon per clock cycle in a random BB84 basis and records the outcomes.
\State In the sifting step, Alice and Bob reveal their bases over a classical channel and retain only matching basis detection events.
\State Alice informs Bob which detections are signal or decoy states.
\State {Bob estimates the gains and errors of the signal and decoy states, and extracts the single photon yield $Y_1$ and error rate $e_1$ to find the secure key rate}.
\State Error correction and privacy amplification are applied to obtain the final secret key.
\end{algorithmic}
\end{algorithm}

\begin{algorithm}[H]
\caption{Heralded Purification Protocol (HP)}\label{alg:alg2}
\begin{algorithmic}[1]

\Require Realistic quantum emitter with photon statistics of $\{P_0, P_1, P_2\}$
\Ensure Ensure a shared secure key between Alice and Bob using highly pure single photons
\State \textbf{Initialization:} Set laser excitation power to $S_3 > 1$ near saturation and configure Alice's heralding single photon detector (SPD) with its detection efficiency of $\eta_{D}$.
\For{each clock cycle defined by the excitation laser's pulse repetition rate}
    \State Alice excites the device with $S_3$.
    \State The photons emitted from the device are propagated in free-space to Alice's beam splitter (BS): one output to the SPD and the other to Alice's encryption unit.
    \If{the SPD clicks, the heralding is successful}
        \State Alice encodes the photon using the BB84 encryption unit and sends the photon to Bob through the free-space quantum channel.
    \Else
        \State Alice discards the pulse, even if Bob received photons.
    \EndIf
\EndFor
\State Bob measures each received photon per clock cycle in a random BB84 basis and records the outcomes.

\State In the sifting step, Alice informs Bob which pulses were heralded. Alice and Bob retains only matching heralded detection events.
\State Bob estimates the gain and error based on the heralded photons.
\State Error correction and privacy amplification are applied to obtain the final secret key.
\end{algorithmic}
\end{algorithm}

\section{Efficiencies Estimation}

To examine the dependence on the collection and detection efficiencies ($\eta_{C},\eta_{D}$ respectively) and on the beam-splitter (BS) as shown in Sec. IV of the main text, we included some modifications to the emission distributions. \\

Given the source's probability distribution, $\{ P_{n}\}$, and some collection efficiency, $\eta_{C}$, one can perform a modification to the emission distribution of the form:
\begin{align}
    P_{2}' &= P_{2} \eta_{C}^2 \\
    P_{1}' &= P_{1} \eta_{C} + P_{2} \cdot 2 \eta_{C} (1-\eta_{C}) \\
    P_{0}' &= 1-P_{1}' - P_{2}'
\end{align}

Here, we do not add the modification to the channel loss, as this loss is fully defined inside Alice's setup and is inaccessible to Bob and Eve.

The consideration of the BS's parameters and of $\eta_{D}$ in the HP scheme is shown in Eq. \ref{eq:HP} in the main text. \\

With the new emission distributions and the model for the channel described in the previous section, the calculated parameters can be inserted to the SKR equations for either the regular BB84 protocol (for HP) or the decoy state protocol (for DTB), given in \cite{Lo2005DecoyDistribution}, to obtain the behaviour for different channel losses. This yields the corresponding MCL as shown in Sec. IV of the manuscript.

\section{Tomography Mappings for QKD Experimental Results}

\begin{figure}[!thb]
    \centering
    \includegraphics[width=1\linewidth]{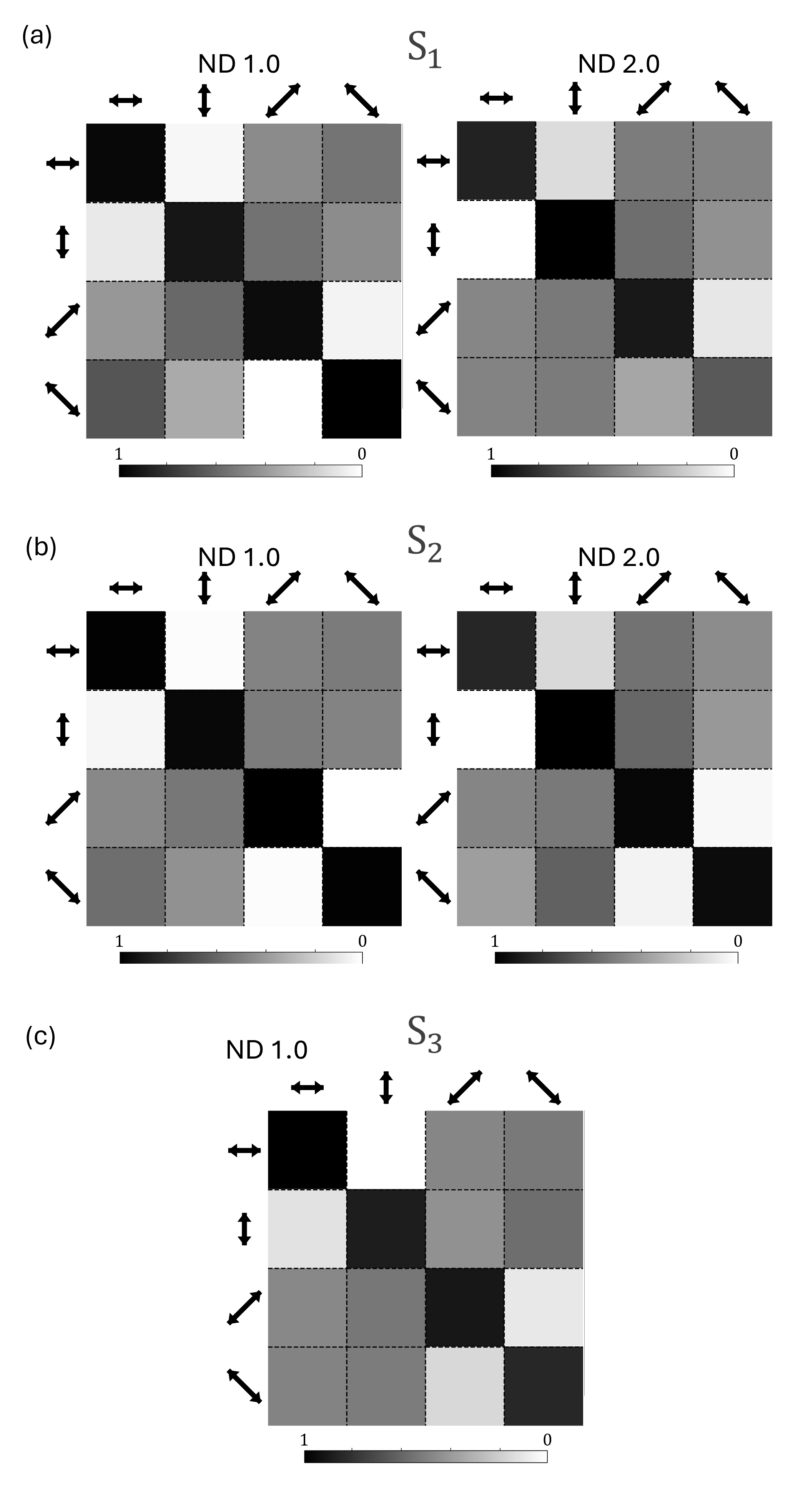}
    \caption{Tomography results for different excitation laser intensities. Intensity corresponding to the decoy state $S_{1}$ (a) and the signal state $S_{2}$ (b) in the DTB protocol, measured with a $1.0$dB ND filter (left) and a $2.0$dB ND filter (right). The intensity used in the HP protocol $S_{3}$ (c), measured with a $1.0$dB ND filter.}
    \label{fig:app3}
\end{figure}

The BB84 experimental demonstration was done by measuring the full Hilbert space of  polarization-based encryption, given a qubit value in a specific basis sent by Alice and a given basis measured by Bob. In Fig. \ref{fig:app3}, the additional results of the tomography maps with an ND filter placed in the quantum channel are presented (for the results without an ND filter, see Fig. \ref{fig:fig6}(b)-(d)). The ND filters are used to demonstrate the channel losses.

\section{Possible values of $g^{(2)}(0)$  with a $\{ P_{0},P_{1},P_{2}\}$ truncated Fock basis}

\begin{figure}[!thb]
    \centering
    \includegraphics[width=1\linewidth]{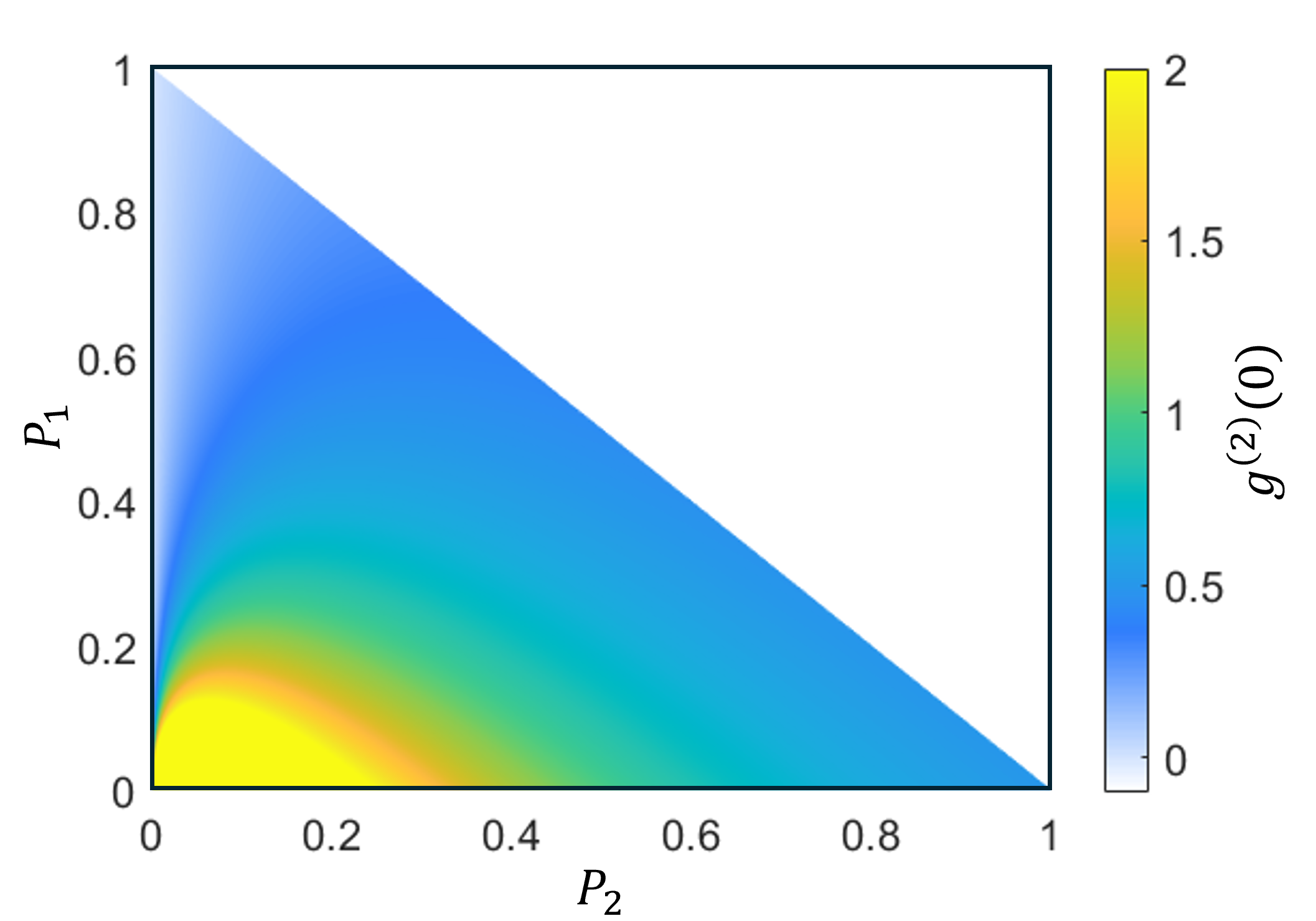}
    \caption{$g^{(2)}(0)$ as a function of $P_{1},\,P_{2}$ for a truncated basis source, showing that values $>0.5$ are possible with increasing $P_{0}$ probability.}
    \label{fig:g2_distribution}
\end{figure}

We provide a derivation that shows that for a realistic imperfect source, which has a total $QY < 1$ and $P_{N\geq3}=0$, the vacuum state affects the second order correlation function such that a valid solution may have $g^{(2)}(0) \geq 0.5$:

First, assume that $P_{0}=0$, meaning there is no vacuum state in the system and $QY=1$. In this case, we use Eqs. \ref{p0new},\ref{eq:g20} to derive:
\begin{equation}
    g^{(2)}(0) = g = \frac{2(1-P_{1})}{(2-P_{1})^{2}}
\end{equation}
Now, we can find the maximum by looking at the derivative:
\begin{align}
    \frac{d g}{dP_{1}} = 0 \rightarrow -2P_{1} = 0 \\ \rightarrow P_{1} = 0 \rightarrow P_{2} = 1 \rightarrow g = \frac{1}{2}
    \nonumber
\end{align}
As expected, a Fock state $\ket{2}$ has a $g^{(2)}(0)$ equaling to exactly $\frac{1}{2}$.

Next, let us assume that $P_{0}\neq 0$. Similarly, we can use Eqs \ref{p0new}, \ref{eq:g20} to derive:
\begin{align} \label{g2_B}
    g^{(2)} = g = \frac{2P_{2}}{(1-P_{0}+P_{2})^{2}}
\end{align}
Where we use $P_{2}$ in our analysis.

Similarly, we can look at the derivative to find:
\begin{align}
    \frac{d g}{dP_{2}} = 0 \rightarrow 1-P_{0}-P_{2} = 0 \\ \rightarrow P_{2} = 1-P_{0}
\end{align}
Assuming that for the maximal value $P_{1} = 0$, we can plug the result into Eq. \ref{g2_B} to find that:
\begin{equation}
    g = \frac{1}{2(1-P_{0})}
\end{equation}
Interestingly, for $P_{0}\rightarrow 0$, we indeed get the previous result of $g = \frac{1}{2}$. But, for $P_{0}\rightarrow 1$, $g\rightarrow\infty$ and the result diverges.

Similar results may be obtained for $P_{1} \neq 0$, as we show in Fig. \ref{fig:g2_distribution}, demonstrating a contour plot of the $g^{(2)}(0)$ results by varying $P_{1}$ and $P_{2}$. In our system, we work with high QYs therefore the extracted probabilities exhibit a $g^{(2)}(0) \sim 0.67$. 

This leads to the conclusion that relatively low vacuum state probabilities and high brightness emission is required for practical implementation of realistic imperfect quantum emitters. In addition, the purity of the source is not only justified by the value of $g^{2}(0)$, but rather can be calculated by the ratio between $\frac{P_{1}}{P_{1}+P_{2}}$ for a $\{ P_{0}, P_{1}, P_{2} \}$ source.\newline

\bibliography{references}

\end{document}